\documentstyle[12pt]{article}

 {\parindent 0.5cm \textwidth 15.0cm \textheight
23.5cm \hoffset=-0.8cm \voffset=-3cm
  \let\LARGE=\large
 \let\large=\normalsize
\newcommand{\be}{\begin{equation}}
\newcommand{\ee}{\end{equation}}
\newcommand{\ba}{\begin{array}{c}}
\newcommand{\ea}{\end{array}}
\newcommand{\dis}{\displaystyle}
\newcommand{\tr}{\hbox{tr}}

\begin{document} \input FEYNMAN
\begin{titlepage} \vspace{0.2in} \begin{flushright}
CPT-93/P.2871 \\ \end{flushright} \vspace*{1.5cm}
\begin{center} {\LARGE \bf  Massive Spin-1 Field Chiral
Lagrangian from \\ an Extended Nambu--Jona-Lasinio Model of QCD\\}
\vspace*{0.8cm}
{\bf Joaquim Prades}\\ \vspace*{1cm} Centre
de Physique Th\'eorique, C.N.R.S. - Luminy, Case 907 \\
F-13288 Marseille Cedex 9, France \\ \vspace*{1.8cm}
{\bf   Abstract  \\ } \end{center} \indent

We present a calculation of the coupling constants in the
massive spin-1 field
chiral Lagrangian to chiral ${\cal O}(p^3)$
in the context of the extended Nambu--Jona-Lasinio
model described in Ref. \cite{bbr}.
Phenomenological applications of this Lagrangian
to anomalous and non-anomalous low-energy hadronic
transitions involving
spin-1 particles are also discussed. \vfill \begin{flushleft}
January 1993 \\
CPT-93/P.2871 \end{flushleft} \end{titlepage}

\section{Introduction} \label{1} \indent

At low-energies the strong, electromagnetic and weak
interactions of the lightest pseudoscalar mesons can be described
by an effective chiral Lagrangian \cite{sw,gl}. This Lagrangian
depends on a number of coupling constants which are not
fixed by symmetry requirements alone but are, in principle,
determined by the dynamics of the underlying QCD theory. Recently,
there have been attempts to derive the low-energy effective
chiral action of QCD \cite{bbr,ert} with and without the
inclusion of the low-lying resonances. The lightest vector,
axial-vector and scalar resonances play an important r\^ole
in the determination of
the low-energy interactions between pseudoscalar mesons in
the context of chiral Lagrangians at ${\cal O} (p^4)$ as
shown in Ref. \cite{egpr}.

In this work we shall study the chiral Lagrangians
for spin-1 particles to ${\cal O} (p^3)$ in the
chiral expansion, in the context of the extended
Nambu--Jona-Lasinio (ENJL) cut-off model described
in Ref. \cite{bbr}. In particular, we shall calculate
the couplings of spin-1 particles to Goldstone bosons
and external sources. As an application, we shall then
predict the decay rates for the transitions
$V \to \pi \gamma$, $V \to \pi \pi \pi$, $V \to \pi \pi
\gamma$, $V \to V' \gamma$, $V \to V' \pi$, $A \to
\pi \pi \pi$, $A \to \pi \gamma$, $A \to V \pi$ and
$A \to V \gamma$. Here $V$ and $A$ denote vector and axial-vector
particles. We shall also discuss the vector meson
dominance predictions for the $\pi^0 \to \gamma \gamma^*$
and $K_L \to \pi^0 \gamma^* \gamma^* \to \pi^0
e^+ e^-$ transitions. The paper is organized as follows.
In Section \ref{2b} we give a brief summary of what is known
about low-energy chiral Lagrangians from general symmetry
requirements alone. In Section
\ref{2} we summarize some of the ENJL results found in Ref.
\cite{bbr}. In Section \ref{3} we shall then study the anomalous
sector of the theory and in Section \ref{4} the non-anomalous
sector. Finally, in Section \ref{5} we discuss the phenomenological
applications to the decays involving spin-1 particles.
The conclusions are given in Section \ref{6}.

\setcounter{equation}{0}
\section{The low-energy chiral Lagrangians} \label{2b}
\indent

We shall give a summary of what is known at present   about
low-energy mesonic Lagrangians, from the chiral invariance
properties of ${\cal L}_{\rm QCD}$ alone.
In the pseudoscalar sector, the terms in the effective Lagrangian
${\cal L}_{\rm eff}$ with the lowest chiral
dimension, {\it i.e.} ${\cal O} (p^2)$, are

\be\label{22}
{\cal L}_{\rm eff}^{(2)} = {1\over 4} \, f^2_0 \, \Big
\{ \tr \,\left( D_\mu U D^\mu \, U^\dagger \right)\, + \,
\tr \, (\chi U^{\dagger} + U \chi^\dagger) \Big \} \ee

\vspace*{0.5cm}

\noindent where $D_{\mu}$ denotes the covariant derivative

\be\label{23}
D_{\mu} U = \partial_\mu U -i( v_\mu + a_\mu)U
+ i U(v_\mu - a_\mu)\,,
\ee

\vspace*{0.5cm}

\noindent and $U \equiv \exp \left( -\frac{\dis \sqrt 2 i
\Phi}{\dis f_0} \right)$ an SU(3) matrix incorporating the
octet of pseudoscalar mesons

\be \label{eta8}
\ba \Phi (x)\,=\,\frac{\dis \vec{\lambda}}{\dis \sqrt 2}
\, \vec{\varphi}
\, = \, \left( \begin{array}{ccc} \frac{\dis \pi^0}{\dis \sqrt 2}
\, + \, \frac{\dis \eta_8}{\dis \sqrt 6}
 & \pi^+ & K^+ \\ \pi^- & - \frac{\dis \pi^0}{\dis \sqrt 2} \,
+ \, \frac{\dis \eta_8} {\dis \sqrt 6}  & K^0 \\ K^- & \overline
{K^0} & - \frac{\dis 2 \, \eta_8}{\dis \sqrt 6} \end{array}
\right) \ea . \ee

\vspace*{0.5cm}

\noindent In Eq. (\ref{23}) $v_\mu$, $a_\mu$ are external
3 $\times$ 3 vector and axial-vector field matrices.
In Eq. (\ref{22})

\be\label{24} \matrix{
\chi \, = \, 2 \, B_0\, (s(x)\, +\, i\, p(x))\,, \hfill \cr\cr}\ee

\vspace*{0.5cm}

\noindent with $s$ and $p$ external scalar and pseudoscalar 3
$\times$ 3 field matrices. In practice

\be\label{mass1} \matrix{
\chi\,=\,2\, B_0\, {\cal M} \hfill\cr\cr}\ee

\vspace*{0.5cm}

\noindent with ${\cal M}$ the 3 $\times$ 3 flavour matrix
${\cal M} \,=\, {\rm diag} (m_u, m_d, m_s)$ which collects the
light-quark masses.
The constants $f_0$ and $B_0$ are not fixed by chiral symmetry
requirements. The constant $f_0$ can be obtained from
$\pi \to \mu \nu$ decay \cite{pdg},

\be \label{25}
f_0 \simeq f_{\pi} \simeq 93  \, {\rm MeV}. \ee

\noindent The constant $B_0$ is related to the vacuum expectation
 value

\be\label{26}
\langle 0 \vert \bar {q} q
\vert 0 \rangle_{\vert q=u,d,s} = \,-\, f_0^2 \, B_0
\, (1\, +\, {\cal O} ({\cal M})).
\ee

\vspace*{0.5cm}

\noindent In the absence of the U(1)$_A$ anomaly, the SU(3) singlet
$\eta_1$ field becomes the ninth Goldstone boson which is
incorporated in the $\Phi(x)$ field as

\be \label{eta1}
\Phi(x) \,=\,\frac{\dis \vec{\lambda}}{\dis \sqrt 2} \,
\vec{\varphi} \,+\, \frac{\dis \eta_1}{\dis \sqrt 3}
\,\mbox{\LARGE\bf 1}\,.
\ee

\vspace*{0.5cm}

The terms in ${\cal L}_{\rm eff}$ of ${\cal O}(p^4)$ are
also known. They have been discussed extensively by Gasser
and Leutwyler \cite{gl}.

We are interested in an effective
 Lagrangian ${\cal L}^R_{\rm eff}$ which also incorporates
chiral couplings of fields of massive $1^-$, $1^+$ and $0^+$
states to
the Goldstone fields. We shall restrict ourselves to the
sector of spin-1 particles.
The general method to construct these couplings was
described a long time ago in Ref. \cite{cwz}.
An explicit construction can be found
in Refs. \cite{egpr} and \cite{eglpr}. As discussed in Ref.
\cite{eglpr},
the choice of fields to describe chiral invariant
couplings involving spin-1 particles is not unique and, when
the vector modes are integrated out, leads to ambiguities in
the context of chiral perturbation theory to ${\cal O} (p^4)$
and higher. As shown in \cite{eglpr}, these ambiguities are,
however, removed when consistency with the short-distance behaviour
of QCD is incorporated. The effective Lagrangian which we shall choose
here to describe spin-1 particle couplings corresponds to the
so-called
model II in Ref. \cite{eglpr}. In this model the linear couplings of
vector and axial-vector
fields to the pseudoscalar mesons and external fields start at
chiral ${\cal O} (p^3)$.
The most general Lagrangian ${\cal L}^R_{\rm eff}$ for spin-1
particles to lowest non-trivial order in the chiral expansion
and linear in the spin-1 fields for the interaction terms
is then obtained by adding to
${\cal L}^{(2)}_{\rm eff}$ in Eq. (\ref{22}) the vector Lagrangian

\be\label{47} \matrix{
{\cal L}_V = & -\, \frac{\dis 1}{\dis 4} \,
\tr\left(V_{\mu \nu} V^{\mu \nu} - 2\, M_V^2 \, V_{\mu}V^{\mu}\right)
\hfill \cr\cr \hphantom{{\cal L}_V =}&
\, - \, \frac{\dis 1}{\dis 2 \sqrt 2} \,
\left[f_V \, \tr\left(V_{\mu \nu} f_{(+)}^{\mu \nu} \right)
+ \, i g_V \, \tr\left(V_{\mu \nu}[\xi^{\mu},\xi^{\nu}]\right)\right]
\hfill \cr\cr \hphantom{{\cal L}_V =}&
+ \, i \, \alpha_V \, \tr\left(V_\mu [\xi_\nu, f_{(-)}^{\mu \nu}]
\right)
\, + \, \beta_V \, \tr\left(V_\mu [\xi^\mu, \chi_{(-)}]\right) \,
+ \, i \, \theta_V \, \epsilon_{\mu \nu \alpha \beta} \,
\tr \left(V^\mu \xi^\nu \xi^\alpha \xi^\beta \right)
\hfill \cr\cr \hphantom{{\cal L}_V =}&
+ \, h_V \, \epsilon_{\mu \nu \alpha \beta} \, \tr \left(
V^\mu \left\{ \xi^\nu, f_{(+)}^{\alpha \beta} \right\} \right)\,,
\hfill \cr} \ee

\vspace*{0.5cm}

\noindent and the axial-vector Lagrangian

\be\label{48} \matrix{
{\cal L}_A = & -\, \frac{\dis 1}{\dis 4} \,
\tr\left(A_{\mu \nu} A^{\mu \nu} -
2M_A^2 A_{\mu}A^{\mu}\right) \, - \, \frac{\dis 1}{\dis 2 \sqrt 2} \,
f_A \, \tr\left( A_{\mu \nu} f_{(-)}^{\mu \nu} \right)
\hfill \cr\cr \hphantom{{\cal L}_A =}&
+ \, i \, \alpha_A \, \tr\left(A_\mu [\xi_\nu, f_{(+)}^{\mu \nu}]
\right)
\, + \, \gamma^{(1)}_A \, \tr \left( A_\mu \xi_\nu \xi^\mu \xi^\nu
\right) \, + \, \gamma^{(2)}_A \, \tr \left( A_\mu \left \{ \xi^\mu,
 \xi^\nu \xi_\nu \right \} \right)
\hfill \cr\cr \hphantom{{\cal L}_A =}&
+ \, \gamma^{(3)}_A \, \tr \left( A_\mu \xi_\nu \right) \, \tr
\left( \xi^\mu \xi^\nu \right) \,
+ \, \gamma^{(4)}_A \, \tr \left( A_\mu \xi^\mu \right) \, \tr
\left( \xi^\nu \xi_\nu \right)
\hfill \cr\cr \hphantom{{\cal L}_A =}&
\, + \,h_A \, \epsilon_{\mu \nu \alpha \beta} \, \tr
\left( A^\mu \left\{ \xi^\nu, f_{(-)}^{\alpha \beta} \right\} \right)
\,. \hfill \cr} \ee

\vspace*{0.5cm}

The notation here is the following.
 We assume nonet symmetry for the spin-1
particles, so that the spin-1 fields we consider are

\be \label{nonet} \matrix{
V^\mu \,=\, V^\mu_8\,+\, V^\mu_1 \hfill\cr\cr
A^\mu \,=\, A^\mu_8\,+\, A^\mu_1
\hfill\cr\cr}\ee

 \vspace*{0.5cm}

\noindent where $V^\mu_8$ and  $A^\mu_8$
are SU(3)$_V$ octets while $V^\mu_1$ and  $A^\mu_1$
are singlets. The vector field matrix $V^{\mu}_8(x)$:

\be \label{w8} \ba V^\mu_8 (x) \equiv
\left( \begin{array}{ccc} \frac{\dis \rho^0}{\dis \sqrt 2}
\, + \, \frac{\dis \omega_8}{\dis \sqrt 6}
 & \rho^+ & K^{* +} \\ \rho^- & - \frac{\dis \rho^0}{\dis \sqrt 2} \,
+ \, \frac{\dis \omega_8} {\dis \sqrt 6}  & K^{* 0} \\ K^{*
-} & \overline {K^{* 0}} & - \frac{\dis 2 \, \omega_8}{\dis \sqrt 6}
\end{array} \right)^\mu \, ,\ea \ee

\vspace*{0.5cm}

\noindent represents the SU(3)$_V$ octet of spin-parity $1^-$-particles
and the axial-vector field matrix $A^{\mu}_8(x)$:

\vspace*{0.5cm}

\be \ba A^\mu_8 (x) \equiv
\left( \begin{array}{ccc} \frac{\dis a_1^0}{\dis \sqrt 2}
\, + \, \frac{\dis (f_1)_8}{\dis \sqrt 6}
 & a_1^+ & K_1^{+} \\ a_1^- & - \frac{\dis a_1^0}{\dis \sqrt 2} \,
+ \, \frac{\dis (f_1)_8} {\dis \sqrt 6}  & K_1^{0} \\ K_1^{-} &
\overline {K_1^{0}} & - \frac{\dis 2 \, (f_1)_8}{\dis \sqrt 6}
\end{array} \right)^\mu \, ,\ea \ee

\vspace*{0.5cm}

\noindent represents the SU(3)$_V$ octet of spin-parity
$1^+$-particles. The vector field matrix $V_1(x)$:

\vspace*{0.5cm}

\be \label{w1} \ba
V_1 (x) \equiv \frac{\dis \omega_1}{\dis \sqrt 3} \,
\mbox{\LARGE\bf 1} \, ,\ea \ee

\vspace*{0.5cm}

\noindent represents
the SU(3)$_V$ singlet of spin-parity $1^-$-particles
and the axial-vector field matrix $A_1(x)$
represents the corresponding
SU(3)$_V$ singlet of spin-parity
$1^+$-particles.

\noindent The vector and axial-vector field strength
tensors are defined as follows

\be\label{39}
V_{\mu\nu}=d_{\mu}V_{\nu}-d_{\nu}V_{\mu}
\hspace*{2cm} {\hbox{and}}\hspace*{2cm}
A_{\mu\nu}=d_{\mu}A_{\nu}-d_{\nu}A_{\mu}\;.
\ee

\vspace*{0.5cm}

\noindent Here, the covariant derivative $d_\mu$
acts on the octet multiplets denoted by
$R_8$ as follows

\be\label{36}
d_{\mu}R_8 = \partial_{\mu}R_8 + [\Gamma_{\mu},R_8]
\ee

\vspace*{0.5cm}

\noindent and trivially on the singlets.
The connection $\Gamma_\mu$ is given by

\be\label{37}
\Gamma_{\mu} = {1 \over 2}\, \left\{ \xi^{\dagger}
\, \left[ \partial_{\mu} - i (v_{\mu}+a_{\mu}) \right]
\, \xi + \xi\, \left[ \partial_{\mu} - i (v_{\mu}-a_{\mu})
\right] \, \xi^{\dagger} \right\} \, .\ee

\vspace*{0.5cm}

\noindent The axial-vector field matrix
$\xi_\mu$ is defined by

\be\label{41}
\xi_{\mu} = i\, \left\{ \xi^{\dagger} \, \left[ \partial_{\mu}
- i (v_{\mu}+a_{\mu}) \right] \, \xi - \xi \, \left[
\partial_{\mu} - i (v_{\mu}-a_{\mu}) \right] \,
\xi^{\dagger} \right\} = i\, \xi^{\dagger} D_{\mu} U
\xi^{\dagger} = \xi_{\mu}^{\dagger} \,. \ee

\vspace*{0.5cm}

\noindent Finally, in Eqs. (\ref{47}) and (\ref{48}),
we have also introduced the following objects

\be\label{43}
\chi_{(\pm)} = \xi^{\dagger}\chi\xi^{\dagger} \pm \xi\chi^{\dagger}
\xi \, ,\ee

\vspace*{0.5cm}

\noindent and

\be\label{44}
f^{\mu\nu}_{(\pm)} = \xi F_L^{\mu \nu} \xi^{\dagger} \pm
\xi^{\dagger} F_R^{\mu \nu} \xi\,,
\ee

\vspace*{0.5cm}

\noindent with $F_L^{\mu \nu}$ and $F_R^{\mu \nu}$ the external
field-strength tensors

\be\label{28}
F_L^{\mu \nu} = \partial^{\mu} l^{\nu} - \partial^{\nu} l^{\mu}
- i \, [l^{\mu}, l^{\nu}] \, ,\ee

\be\label{29}
F_R^{\mu \nu} = \partial^{\mu} r^{\nu} - \partial^{\nu} r^{\mu}
- i \, [r^{\mu},r^{\nu}]
\ee

\vspace*{0.5cm}

\noindent associated with
the external left ($l_{\mu}$) and right ($r_{\mu}$)
field sources

\be\label{30}
l_{\mu} = v_{\mu} - a_{\mu}{\,\,\,},\;\;\;r_{\mu} =
v_{\mu} + a_{\mu}\, .\ee

\vspace*{0.5cm}

In this paper,
we are also interested in couplings of vector and axial-vector
quadratic in the spin-1 fields. The complete set of these quadratic
interaction terms for vector meson fields,
to lowest chiral ${\cal O} (p^2)$, is

\be \label{49} \matrix{
{\cal L}_{VV}^{(2)} = &
\frac{\dis 1}{\dis 2} \,
\delta^{(1)}_V \, \tr \left(\left[ \xi_\mu, \xi_\nu \right]
\left[ V^\mu, V^\nu \right] \right) \, + \,
\frac{\dis 1}{\dis 2} \,\delta^{(2)}_V \,
\tr \left( \left[ V_\mu, \xi_\nu \right] \left[ V^\nu,
\xi^\mu \right] \right)
\hfill \cr\cr \hphantom{{\cal L}_V^{(2)} =}& + \,
\frac{\dis 1}{\dis 2} \, \delta^{(3)}_V \, \tr \left( \left[
V_\mu, \xi^\mu \right] \left[ V_\nu, \xi^\nu \right] \right)
\, + \, \frac{\dis 1}{\dis 2} \,
\delta^{(4)}_V \, \tr \left( V^\mu V^\nu \xi_\mu \xi_\nu \right)
\, + \, \frac{\dis 1}{\dis 2} \,
\delta^{(5)}_V \, \tr \left( V^\mu V_\mu \xi^\nu \xi_\nu \right)
\hfill \cr\cr \hphantom{{\cal L}_V^{(2)} =}&
\, + \, \frac{\dis 1}{\dis 2} \,
\delta^{(6)}_V \, \tr \left( V^\mu \xi^\nu V_\mu \xi_\nu \right)
\, + \,\frac{\dis 1}{\dis 2} \,
i\, \phi_V \, \tr \left( V_\mu \left[ V_\nu , f_{(+)}^{\mu \nu}
\right] \right) \hfill \cr\cr \hphantom{{\cal L}_V^{(2)} =}&
\, + \, \frac{\dis 1}{\dis 2} \,
\sigma_V \, \epsilon_{\mu \nu \alpha \beta}
\, \tr \left( V^\mu \left\{ \xi^\nu, V^{\alpha \beta} \right\}
\right) \,. \hfill \cr\cr}  \ee

\vspace*{0.5cm}

\noindent The corresponding interaction terms
for axial-vector meson fields are

\be \label{50} \matrix{
{\cal L}_{AA}^{(2)} = &
\frac{\dis 1}{\dis 2} \,
\delta^{(1)}_A \, \tr \left(\left[ \xi_\mu, \xi_\nu \right]
\left[ A^\mu, A^\nu \right] \right) + \, \frac{\dis 1}{\dis 2} \,
\delta^{(2)}_A \,
\tr \left( \left[ A_\mu, \xi_\nu \right] \left[ A^\nu,
\xi^\mu \right] \right)
\hfill \cr\cr \hphantom{{\cal L}_V^{(2)} =}&
\, + \, \frac{\dis 1}{\dis 2} \,
\delta^{(3)}_A \, \tr \left( \left[ A_\mu, \xi^\mu \right]
\left[ A_\nu, \xi^\nu \right] \right)
\, + \, \frac{\dis 1}{\dis 2} \,
\delta^{(4)}_A \, \tr \left( A^\mu A^\nu \xi_\mu \xi_\nu \right)
\, + \, \frac{\dis 1}{\dis 2} \,
\delta^{(5)}_A \, \tr \left( A^\mu A_\mu \xi^\nu \xi_\nu \right)
\hfill \cr\cr \hphantom{{\cal L}_V^{(2)} =}&
\, + \, \frac{\dis 1}{\dis 2} \,
\delta^{(6)}_A \, \tr \left( A^\mu \xi^\nu A_\mu \xi_\nu \right)
\, + \, \frac{\dis 1}{\dis 2} \,i\,
\phi_A \, \tr \left( A_\mu \left[ A_\nu , f_{(+)}^{\mu \nu}
\right] \right) \hfill\cr\cr \hphantom{{\cal L}_A^{(2)} =}&
\, + \, \frac{\dis 1}{\dis 2} \,
\sigma_A \, \epsilon_{\mu \nu \alpha \beta}
\, \tr \left( A^\mu \left\{ \xi^\nu, A^{\alpha \beta} \right\} \right)
\,. \hfill \cr }  \ee

\vspace*{0.5cm}

\noindent At ${\cal O} (p^2)$ there are also terms
which mix axial-vector and vector fields.
The corresponding interaction
Lagrangian has the following form,

\be \label{51} \matrix{
{\cal L}_{VA}^{(2)} =&
i \, A^{(1)} \, \tr \left( V_\mu \left[ A_\nu , f_{(-)}^{\mu \nu}
\right] \right) \, + \, i \, A^{(2)} \, \tr \left( V_\mu
\left[\xi_\nu, A^{\mu \nu} \right] \right) \, + \, i \, A^{(3)} \, \tr
\left( A_\mu \left[ \xi_\nu, V^{\mu \nu} \right] \right)
\hfill\cr\cr \hphantom{{\cal L}_{VA}^{(2)} =}&
\, + \, B \, \tr \left( V_\mu \left[ A^\mu, \chi_{(-)} \right]
\right) \, + \, H \, \epsilon_{\mu \nu \alpha \beta} \, \tr
\left( V^\mu \left\{ A^\nu , f^{\alpha \beta}_{(+)} \right\} \right)
\hfill\cr\cr \hphantom{{\cal L}_{VA}^{(2)} =}&
\, + \, i \, Z^{(1)} \, \epsilon_{\mu \nu \alpha \beta} \, \tr \left(
\xi^\mu \xi^\nu \left\{ A^\alpha, V^\beta \right\} \right) \,
+ i \, Z^{(2)} \, \epsilon_{\mu \nu \alpha \beta} \, \tr \left(
\xi^\mu A^\nu \xi^\alpha V^\beta \right) \,. \hfill \cr}  \ee

\vspace*{0.5cm}

In the present work we shall concentrate on  processes involving
spin-1 particles and we shall not consider possible spin-1 particle
decays to scalar particles.
 Whenever scalar particles can contribute as
intermediate resonances to these transitions, in particular
for $a_1$-decays, we cannot integrate them out since their
masses are lower than those of the axial-vector particles
and they have to be taken into account as propagating particles
described by the corresponding Lagrangian. The effective
Lagrangian for scalar particles that may contribute to the
transitions we are interested in is of chiral ${\cal O}
(p^2)$. To this order, the most general Lagrangian for
$0^+$-particles involving at most one scalar
and two spin-1 fields is

\be \label{52} \matrix{
{\cal L}_{S} =& \frac{\dis 1}{\dis 2} \,
\tr \left( d_\mu S \, d^\mu S
\,-\,M_S^2 \, S^2 \right)
\,+\,c_m \, \tr \left(S\, \chi^+
\right)\,+\,c_d\,\tr\left(S\xi^\mu\xi_\mu\right)\hfill\cr\cr
\hphantom{{\cal L}_{S} =} &
+\,C^{(1)}\,\tr \left(S\left\{A_\mu,\xi^\mu\right\}\right)\,+\,
\frac{\dis 1}{\dis 2}\,C^{(2)}\,\tr\left(S A^\mu A_\mu\right)
\,+\,\frac{\dis 1}{\dis 2}\,D\,\tr\left(S V^\mu V_\mu\right)
 \,. \hfill \cr}  \ee

\vspace*{0.5cm}

\noindent Here, the 3 $\times$ 3 field matrix $S$ represents
the SU(3)$_V$ nonet of scalar fields.
The chiral ${\cal O}(p^2)$ couplings
$c_m$ and $c_d$ in the ENJL cut-off model
have been calculated in Ref. \cite{bbr}.

\setcounter{equation}{0}
\section{The ENJL cut-off model of QCD} \label{2} \indent

Reference \cite {bbr} gives
a systematic study of the low-energy
effective action of the extended Nambu--Jona-Lasinio model
which, at intermediate energies below or of the order of a cut-off
scale $\Lambda_\chi$, is expected to be a good effective realization
of the standard QCD Lagrangian ${\cal L}_{\rm QCD}$.
Here we give a brief summary of the results found there.
The Lagrangian in question is

\be \label{qcd} \ba
{\cal L}_{\rm QCD} \rightarrow
{\cal L}^{\Lambda_\chi}_{\rm QCD} \, + \,
{\cal L}^{\rm S,P}_{\rm NJL}
\, + \, {\cal L}^{\rm V,A}_{\rm NJL} \, + \, {\cal O} \,
\left( \frac{\dis 1} {\dis \Lambda^4_\chi} \right)\,, \cr \cr
{\rm with} \hspace*{0.5cm} {\cal L}^{\rm S,P}_{{\rm NJL}}
\, =\,
\frac{\dis 8 \pi^2 \, G_{\rm S} (\Lambda_\chi)}
{\dis N_c \, \Lambda_\chi^2} \, {\dis \sum_{i, j}}
\, (\bar q^i_R q^j_L)(\bar q^j_L q^i_R)
\cr \cr
{\rm and} \hfill \cr \cr {\cal L}^{\rm V,A}_{{\rm NJL}}
\, =\,
-\, \frac{\dis 8 \pi^2 \, G_{\rm V} (\Lambda_\chi)}
{\dis N_c \, \Lambda_\chi^2} \, {\dis \sum_{i, j}}
\, \left[ (\bar q^i_L \gamma^\mu q^j_L)
(\bar q^j_L \gamma_\mu q^i_L)
\, + \, (L \to R) \right] .
\ea \ee

\vspace*{0.5cm}

\noindent Where $i, j$ are
flavour indices and $\Psi^L_R \equiv
\frac{\dis 1}{\dis 2} \, (1 \pm \gamma_5) \, \Psi$. The
couplings $G_{\rm S}$ and $G_{\rm V}$ are dimensionless
and ${\cal O} (1)$ in the $1/N_c$-expansion.
In the mean-field approximation, these
${\cal L}^{\rm S,P,V,A}_{\rm NJL}$
above are equivalent to the  constituent chiral
quark-mass term \cite{ert}:

\be \label{chb} \ba - \, M_Q (\bar q_R \, U \,
q_L \, + \, \bar q_L \, U^\dagger \, q_R ) \, . \ea \ee

\vspace*{0.5cm}

\noindent In the presence of  this term
it is useful to introduce new
quark fields $Q_L$ and $Q_R$, which we call
``rotated basis'', or constituent, chiral-quarks
 \cite{mg}, defined as follows

\be \label{rotbas} \begin{array}{cc}
Q_L \, = \, \xi \, q_L \, ;& \overline Q_L \, = \, \bar
q_L \, \xi^\dagger \, ;\\ Q_R \, = \, \xi^\dagger \,
 q_R \, ;& \overline  Q_R \, = \, \bar q_R \, \xi \, .
\ea \ee

\vspace*{0.5cm}

\noindent In the rest of this Section we shall
work in  Euclidean space for convenience,
and we shall adopt the
conventions given in Ref. \cite{bbr}.
The effective action, in the presence of external sources
$l_\mu$ and $r_\mu$ and light-quark-mass matrix ${\cal M}$,
can be written as \cite{bbr}

\be \label{53} \matrix{
e^{\Gamma_{\rm eff} \left(H, \xi, W_\mu^{\pm}; l_\mu,
r_\mu, {\cal M}\right)} \,=\, \hfill \cr\cr
\exp \left( -\, {\dis \int} \, {\rm d}^4 x \,
\left\{ \frac{\dis N_c \Lambda_\chi^2}{\dis 8 \pi^2
 G_S (\Lambda_\chi)} \, \tr \, H^2 \,+\,
\frac{\dis N_c \Lambda_\chi^2}{\dis 64 \pi^2 G_V
(\Lambda_\chi)} \, \tr \left( W_\mu^+ W_\mu^+ \,+\,
W_\mu^- W_\mu^- \right) \right\} \right) \hfill \cr\cr
\hspace*{7cm} \times \, \frac{\dis 1}{\dis Z} \,
{\dis \int} \left[ {\cal D} \, G_\mu \right] \, \exp
\Gamma_E \left( {\cal A}_\mu, M \right)
\,, \hfill \cr \cr }
\ee

\vspace*{0.5cm}

\noindent where $W_\mu^\pm(x)$ and $H(x)$ are 3 $\times$
3 auxiliary Hermitian field matrices which under the chiral
 group transform as

\be\label{54} \matrix{
W_\mu^\pm \to h (\Phi,g_{L,R})
\, W_\mu^\pm \, h^{\dagger} (\Phi,g_{L,R})
\,, \hfill \cr\cr
H \to h (\Phi,g_{L,R}) \, H \, h^\dagger (\Phi,g_{L,R})
\, , \hfill \cr\cr} \ee

\noindent where $h(\Phi,g_{L,R})\equiv h$
 is the compensating SU(3)$_V$ transformation
 which appears under the action of the chiral
group $G\equiv$ SU(3)$_L$ $\times$ SU(3)$_R$
 on the coset representative $\xi(\Phi)$ of the
$G$/SU(3)$_V$ manifold, {\it i.e.}

\be\label{31}
\xi(\Phi) \to g_R \, \xi(\Phi) \, h^{\dagger}
\, = \, h \, \xi(\Phi) \, g_L^{\dagger}\;,
\ee

\vspace*{0.5cm}

\noindent where $\xi(\Phi) \xi(\Phi) \,= \,U$
in the chosen gauge.
In the mean-field approximation these field
matrices are replaced by

\be \label{mean} \matrix{
H \Rightarrow M_Q\,\mbox{\LARGE \bf 1}\,,\hfill\cr\cr
W^\pm_\mu \Rightarrow 0\,. \hfill\cr\cr} \ee

\noindent In Eq. (\ref{53}) we have used the short-hand
notation

\be \left[ {\cal D} \, G_\mu \right] \equiv
{\cal D} \, G_\mu \, \exp \left(
-\, \frac{\dis 1}{\dis 4} \,
{\dis \sum^{N_c^2-1}_{a=1}} \, G_{\rho \nu}^{(a)}
G_{\rho \nu}^{(a)} \right) \,  \ee

\vspace*{0.5cm}

\noindent with $G^{(a)}_{\mu \nu}$ the gluon
field strength tensor and

\be \label{effqcd} \ba
\exp \Gamma_E \, ({\cal A}_\mu, M) \, =\,
{\dis \int} \, {\cal D} \, \overline Q \,
{\cal D} \, Q \, \exp {\dis \int} \, {\rm d}^4 x
\, \overline Q D_E Q \, = \, {\rm det} \, D_E\,
\ea \ee

\vspace*{0.5cm}

\noindent where $D_E$ is the Euclidean Dirac operator

\be \label{dirac} \renewcommand{\arraystretch}{1.5}
\ba D_E \, = \, \gamma_\mu \nabla_\mu \, + \, M
\, = \, \gamma_\mu (\partial_\mu
\, + \, {\cal A}_\mu) \, + \, M
\ea \renewcommand{\arraystretch}{1} \ee

\noindent with

\be \label{dcov} \matrix{
{\cal A}_\mu \, = \, i \, G_\mu \, + \, \Gamma_\mu \,
- \frac{\dis i}{\dis 2} \gamma_5 \,\left( \xi_\mu
\,-\, W_\mu^{-} \right) \,-\, \frac{\dis i}{\dis 2}
\, W^+_\mu \, ; \hfill \cr\cr
M \, = \,-\, H \,-\,
\frac{\dis 1}{\dis 2} \, \left( \Sigma \,-\, \gamma_5 \,
\Delta \right) \, \hfill \cr\cr } \ee

\noindent and

\be \label{masa} \matrix{
\Sigma \,=\, \xi^\dagger \, {\cal M} \,\xi^\dagger \,+\,
\xi \, {\cal M}^\dagger \, \xi ; \hfill \cr\cr
\Delta \,=\, \xi^\dagger \, {\cal M}\, \xi^\dagger \,-\,
\xi \, {\cal M}^\dagger \, \xi \,. \hfill \cr\cr
} \ee

\noindent
In Eq. (\ref{dcov}) $G_\mu$ is the gluon field matrix in the
fundamental SU($N_c$) representation and
the connection $\Gamma_\mu$ and axial-vector
field $\xi_\mu$ are given in Eqs. (\ref{37}) and (\ref{41}).
In our calculation of the fermionic determinant
we shall disregard the gluonic corrections due to fluctuations
below the cut-off scale $\Lambda_\chi$ and shall consider a model
that is more like the first alternative envisaged in Ref. \cite{bbr},
where the formal integration over gluon fields has been done
in the path-integral of the generating functional for
Green functions. Our choice here is motivated by the fact that
the results obtained within this alternative in Ref.
\cite{bbr} for the couplings of the low-energy chiral
Lagrangian to ${\cal O} (p^4)$ are not qualitatively
different from those obtained in the same Reference
when the effect of long-distance gluonic interactions
was taken into account.

In Ref. \cite{bbr} it was pointed out that the
effective action in Eq. (\ref{53}) obeys the following
formal symmetry:
the effective action $\Gamma_E ({\cal A}_\mu, M)$
with ${\cal A}_\mu$ and $M$ given in Eq. (\ref{dcov})
can be written formally as the mean-field
approximation expression corresponding to external
sources $l_\mu$ and $r_\mu$ and light-quark-mass matrix
${\cal M}$ redefined as follows,

\be \label{92} \matrix{
l_\mu \to l'_\mu \, = \, l_\mu \, + \,\frac{\dis 1}{\dis 2}
\, \xi^\dagger
\left( W_\mu^+ \,+\, W_\mu^- \right) \xi\,, \hfill \cr\cr
r_\mu \to r'_\mu \, = \, r_\mu \, + \, \frac{\dis 1}{\dis 2}
\, \xi \left( W_\mu^+ \,-\, W_\mu^- \right) \xi^\dagger\,,
\hfill\cr\cr
{\cal M} \to {\cal M}'\,=\, {\cal M}\,+\, \xi\, \sigma
\, \xi\,. \hfill \cr\cr } \ee

\noindent Here,

\be \sigma \equiv H \,-\, M_Q \,
\mbox{\LARGE \bf 1} \ee

\noindent is a $0^+$-field matrix.
We shall use this formal symmetry in the following
Sections.

Once the quarks and gluons in the
effective action in Eq. (\ref{53}) are integrated out
(see Ref. \cite{bbr}), one can get the correct kinetic term
for spin-1 particles after two steps.
The first one is the diagonalization of the quadratic form
in $\xi_\mu$ and $W_\mu^{(-)}$ that defines the
constant $g_A$, {\it i.e.}

\vspace*{0.5cm}

\be \label{ga}
W_\mu^{(-)} \to \widehat W_\mu^{(-)} \, + \,
(1 \, - \, g_A) \, \xi_\mu .
\ee

\vspace*{0.5cm}

\noindent The second one is a scale redefinition of the fields
$W_\mu^{(+)}$ and $\widehat W_\mu^{(-)}$, {\it i.e.}

\vspace*{0.5cm}

\be \label{scale}
V_\mu \, = \, \frac{\dis f_V}{\dis \sqrt 2} \, W_\mu^{(+)}\,,
\hspace*{1cm} A_\mu \, = \, \frac{\dis f_A}{\dis g_A \, \sqrt 2}
 \, \widehat W_\mu^{(-)} \ee

\vspace*{0.5cm}

\noindent that introduces the
physical vector and axial-vector fields $V_\mu$ and $A_\mu$.
All the couplings of the low-energy effective action
can then be obtained as functions of three parameters only,
$g_A$, $M_Q$ and $\Lambda_\chi$.
The ${\cal O}(p^3)$ couplings $f_V$, $g_V$ and $f_A$ have been
already
calculated in Ref. \cite{bbr} with the results to
leading ${\cal O}(N_c)$,

\be \label{fvc} \matrix{
f_V^2 \, = \, \frac{\dis N_c}{\dis 16 \pi^2} \,
\frac{\dis 2}{\dis 3} \, \Gamma (0,x)\,, \hspace*{2cm}
f_A^2 \, = \, \frac{\dis N_c}{\dis 16 \pi^2} \,
\frac{\dis 2 \, g_A^2}{\dis 3} \left[ \Gamma (0,x) \, - \,
\Gamma(1,x) \right]\, , \hfill \cr \cr
g_V \, = \, \frac{\dis N_c}{\dis 16 \pi^2} \, \frac{\dis 1}{\dis
3 \, f_V} \, \left[ (1 - g_A^2) \, \Gamma(0,x) \, + \, 2 \, g_A^2
\, \Gamma(1,x) \right] \,.\hfill\cr\cr }
\ee

\vspace*{0.5cm}

\noindent Here,

\be \label{xd}
x \, = \, \frac{\dis M_Q^2}{\dis \Lambda_\chi^2} \ee

\vspace*{0.5cm}

\noindent and the function $\Gamma(n,x)$ is
the incomplete Gamma function

\be \label{gamma} \matrix{
\Gamma(n,x) \, = \, {\dis \int^\infty_x}
\frac{\dis {\rm d} z}{\dis z} \, e^{-z} \, z^n
 , \hspace*{1cm}  n \, = \, 0, \, 1, \ldots
\hfill \cr\cr } \ee

\vspace*{0.5cm}

The correct kinetic term for $0^+$-particles is
obtained after a scale redefinition of the $\sigma$
field in Eq. (\ref{92}), {\it i.e.}

\be \label{scal} \matrix{
S \,=\, \lambda_S \, \sigma \hfill\cr\cr} \ee

\noindent that introduces the physical $0^+$-field
$S$. The constant $\lambda_S$ was calculated
in this model in Ref. \cite{bbr},

\be \label{lambdas} \matrix{
\lambda_S^2\,=\,\frac{\dis N_c}{\dis 16 \pi^2}\,
\frac{\dis 2}{\dis 3} \, \left[3\, \Gamma(0,x)\,-\,
2\, \Gamma(1,x)\right]\,. \hfill\cr\cr}\ee

\vspace*{0.5cm}

In the present work we want to calculate the rest of the ${\cal O}
(p^3)$ couplings in the Lagrangians in Eqs. (\ref{47}),
(\ref{48})  and (\ref{52}) as well as
the ${\cal O} (p^2)$ couplings of the interaction Lagrangians in
Eqs. (\ref{49}), (\ref{50}) and (\ref{51}).
We shall later study some of the
phenomenological applications of our results.
We first consider the terms that carry a Levi-Civita
pseudotensor, {\it i.e.} the abnormal intrinsic parity terms
\cite{bij2}.

\setcounter{equation}{0}
\section{The anomalous sector} \label{3} \indent

Here we are concerned with the imaginary part of the effective
action.
The lowest order Lagrangian in the abnormal intrinsic parity
sector describing the interaction of pseudoscalar mesons and
external $l_\mu$ and $r_\mu$ sources is the one given by the
Wess-Zumino (WZ) effective action \cite{wz,ew}.
The ${\cal O} (p^6)$ imaginary part of the
effective action has been obtained in Ref. \cite{bij2} in
the mean-field approximation of the model
we consider here. Using the formal symmetry
in the ENJL cut-off model described above in Section \ref{2},
we can obtain to ${\cal O} (p^3)$ the abnormal
intrinsic parity interaction
terms for spin-1 particles from an effective
action that formally has the same expression as that of
the WZ action but with
the external sources $l_\mu$ and $r_\mu$ replaced by the primed
sources in Eq. (\ref{92}).

The WZ effective action \cite{wz,ew}
has the following explicit form in a scheme
where vector currents are conserved:

\be \label{wz} \matrix{
S_{WZ} \left[ U, l, r \right] = & - \frac{\dis i \, N_c}
{\dis 240 \pi^2} \, {\dis \int} \, d \sigma^{ijklm} \, \tr \left(
\Sigma_i^L \, \Sigma_j^L \, \Sigma_k^L \, \Sigma_l^L \, \Sigma_m^L
\right) \hfill \cr \cr \cr
 \hphantom{S \left[ U, l, r \right]_{WZ} =} &
- \, \frac{\dis i \, N_c}{\dis 48 \pi^2} \, {\dis \int} \, d^4 x \,
\epsilon_{\mu \nu \alpha \beta} \, \left( W ( U, l, r)
^{\mu \nu \alpha \beta} \, - \, W ( \mbox{\LARGE \bf 1}, l, r)
^{\mu \nu \alpha \beta} \right) \,,\hfill \cr \cr} \ee

\vspace*{0.5cm}
\noindent where

\be \matrix{ W ( U, l, r)_{\mu \nu \alpha \beta} = &
\tr \left( U \, l_\mu \, l_\nu \, l_\alpha \, U^\dagger \, r_\beta \,
+ \, \frac{\dis 1}{\dis 4} \, U \, l_\mu \, U^\dagger \, r_\nu \,
U \, l_\alpha \, U^\dagger \, r_\beta \, + \, i \, U \, \partial_\mu
\, l_\nu \, l_\alpha \, U^\dagger \, r_\beta \hfill \right. \cr \cr
\hphantom{W ( U, l_\mu, r_\mu)_{\mu \nu \alpha \beta} = } &
+ \, i \, \partial_\mu \, r_\nu \, U \, l_\alpha \, U^\dagger \,
r_\beta \, - \, i \, \Sigma_\mu^L \, l_\nu \, U^\dagger \, r_\alpha \,
U \, l_\beta \, + \, \Sigma_\mu^L \, U^\dagger \partial_\nu \,
r_\alpha \, U \, l_\beta \hfill \cr \cr
\hphantom{W ( U, l_\mu, r_\mu)_{\mu \nu \alpha \beta} = } &
- \, \Sigma^L_\mu \, \Sigma^L_\nu \, U^\dagger \, r_\alpha \, U \,
l_\beta \, + \, \Sigma^L_\mu \, l_\nu \, \partial_\alpha \, l_\beta
\, + \, \Sigma^L_\mu \, \partial_\nu \, l_\alpha \, l_\beta
\hfill \cr \cr
\hphantom{W ( U, l_\mu, r_\mu)_{\mu \nu \alpha \beta} = } &
\left. - \, i \, \Sigma^L_\mu \, l_\nu \, l_\alpha \, l_\beta \, + \,
\frac{\dis 1}{\dis 2} \, \Sigma^L_\mu \, l_\nu \, \Sigma^L_\alpha \,
l_\beta \, - \, i \, \Sigma^L_\mu \, \Sigma^L_\nu \, \Sigma^L_\alpha
\, l_\beta \right) \hfill \cr \cr
\hphantom{W ( U, l_\mu, r_\mu)_{\mu \nu \alpha \beta} = } &
- \, \left( L \leftrightarrow R \right) \hfill \cr \cr
{\rm with}& \Sigma^L_\mu = U^\dagger \, \partial_\mu \, U\,,
\hspace*{1cm} \Sigma^R_\mu = U \, \partial_\mu \, U^\dagger
\hspace*{0.5cm} {\rm and} \hspace*{0.5cm} \epsilon_{0123} = 1\,.
\hfill \cr \cr } \ee

\noindent Here, $\left( L \leftrightarrow R \right)$ stands for the
interchanges

\be \label{inter}
U \leftrightarrow U^\dagger\,, \hspace*{0.5cm} l_\mu \leftrightarrow
r_\mu\,, \hspace*{0.5cm} \Sigma^L_\mu \leftrightarrow \Sigma^R_\mu\,.
\ee

\vspace*{0.5cm}
\noindent The first term in Eq. (\ref{wz}) is an integration
of an antisymmetric SU(3)$_L$ $\times$ SU(3)$_R$ invariant
fifth-rank tensor that does not contain external sources
over a five-dimensional sphere whose boundary is four-dimensional
Minkowski space \cite{ew}.

We can now show how the formal symmetry of the ENJL cut-off
version of QCD explained in Section \ref{2} works.
As an example we are going to calculate the term that
is modulated by the coupling constant $h_V$ in Eq.
(\ref{47}). We can write down
the interaction Lagrangian linear in the $W^{(+)}_\mu$
field as

\be \label{exan}
{\cal L}_I^{W^{(+)}} \equiv \tr \, \left(
W^{(+)}_\mu J^\mu \right)\,.
\ee

\vspace*{0.5cm}

\noindent To obtain the current $J^\mu$, first we
calculate the left ($L_\mu$) and right ($R_\mu$)
anomalous chiral currents from the WZ effective
action above, {\it i.e.}

\be \label{leftr} \matrix{
\widetilde L^\mu \, \equiv \,
\frac{\dis \delta S_{WZ}}{\dis \delta l_\mu}
\equiv L^\mu \, +\,& {\rm non-chirally \,\, covariant
\,\, polynomial} \hfill\cr
\hphantom{\widetilde L^\mu \, \equiv \,
\frac{\dis \delta S_{WZ}}{\dis \delta l_\mu} \equiv
L^\mu \, +\,} &{\rm in \,\, external \,\,sources}\, ,
\hfill\cr\cr \widetilde R^\mu \, \equiv \,
\frac{\dis \delta S_{WZ}}{\dis \delta r_\mu}
\equiv R^\mu \, +\,& {\rm non-chirally \,\, covariant
\,\, polynomial} \hfill\cr \hphantom{
\widetilde R^\mu \, \equiv \,
\frac{\dis \delta S_{WZ}}{\dis \delta r_\mu}
\equiv R^\mu \, +\,}&{\rm in \,\, external \,\,sources}\,
. \hfill\cr\cr}\ee

\noindent The anomalous currents $\widetilde L^\mu$ and
$\widetilde R^\mu$  have the well known structure
\cite{bz,le} which consists of a chirally covariant
part ($L^\mu$ and $R^\mu$) that is scheme independent  and
a non-chirally covariant polynomial in external sources
that depends on the scheme. Of course, the physics is contained
in the chirally covariant part and does not change
with the scheme. It is then easy to see that the current
 $J^\mu$ we want in Eq. (\ref{exan})
can be obtained as follows,

\be \label{symm}
J^\mu\,=\, \frac{\dis 1}{\dis 2}\, \left[ \xi L^\mu
\xi^\dagger \,+\, \xi^\dagger R^\mu
\xi \right]_{\left| \begin{array}{c}
 l_\mu \to l_\mu + \frac{1}{2} \xi^\dagger
W_\mu^{(-)} \xi \\ r_\mu \to r_\mu - \frac{1}{2}
\xi W_\mu^{(-)} \xi^\dagger \end{array} \right.} \,. \ee

\vspace*{0.5cm}

\noindent The result we get for the term we are interested in
is

\be \label{res}
{\cal L}_I^{W^{(+)}} \doteq \frac{\dis N_c}{\dis 48 \pi^2}
\, \frac{\dis 1}{\dis 4} \,
\epsilon^{\mu \nu \alpha \beta} \, \tr \, \left(
W^{(+)}_\mu \left\{ 3 \, \xi_\nu \,-\, 2\, W^{(-)}_\nu
, f^{(+)}_{\alpha \beta} \right\} \right) \,.
\ee

\vspace*{0.5cm}

\noindent As was explained in Section \ref{2},
to obtain the interaction Lagrangian of the
physical vector field $V_\mu$ from ${\cal L}^{W^{(+)}}_I$
we have to perform the shift in Eq. (\ref{ga}) and the scale
redefinition in Eq. (\ref{scale}). These two operations
introduce the couplings $g_A$, $f_V$ and $f_A$ which collect the
information of the underlying theory that has been integrated
out. The Lagrangian that we obtain then for the term we are
considering is

\be \label{resf}
{\cal L}_I^V \doteq \frac{\dis N_c}{\dis 16 \pi^2}
\, \frac{\dis \sqrt 2}{\dis 4 f_V} \, \left(1\,-\,
\frac{\dis 2}{\dis 3} \, (1\,-\,g_A) \right) \,
\epsilon^{\mu \nu \alpha \beta} \, \tr \, \left(
V_\mu \left\{ \xi_\nu, f^{(+)}_{\alpha \beta} \right\}
\right) \,. \ee

\vspace*{0.5cm}

After performing the same kind of calculations as in the
example above, for the other abnormal intrinsic
parity terms in
Eqs. (\ref{47}), (\ref{48}), (\ref{49}),
(\ref{50}) and (\ref{51}) we get the following
results for these couplings:

\be \label{coup} \matrix{
\theta_V = \frac{\dis N_c}{\dis 16 \pi^2} \,
\frac{\dis \sqrt 2}{\dis 3 \, f_V} \, \left( 1 +
\frac{\dis \left(1 + g_A \right)
\left(1 - g_A^2 \right) }{\dis 8} \right)\,, &
h_V  =  \frac{\dis N_c}{\dis 16 \pi^2} \,
\frac{\dis \sqrt 2}{\dis 4 \, f_V} \, \left( 1 -
\frac{\dis 2}{\dis 3} \, \left( 1-g_A \right) \right)\,,
\hfill \cr \cr h_A \, = \, \frac{\dis N_c}{\dis 16 \pi^2} \,
\frac{\dis g_A^2 \, \sqrt 2}{\dis 12 \, f_A} \,, &
\sigma_V \, = \, \frac{\dis N_c}{\dis 16 \pi^2} \, \frac{\dis
1}{\dis 2 \, f_V^2} \, \left( 1 - \frac{\dis 2}{\dis 3}
\left( 1 - g_A \right) \right)\,, \hfill \cr \cr
\sigma_A \, = \, \frac{\dis N_c}{\dis 16 \pi^2} \,
\frac{\dis g_A^2}{\dis 6 \, f_A^2} \,, &
H \, = \, - \, \frac{\dis N_c}{\dis 16 \pi^2} \, \frac{\dis
g_A}{\dis 6 \, f_A \, f_V}\,, \hfill \cr \cr
Z^{(1)} \, = \, - \frac{\dis N_c}{\dis 16 \pi^2} \,
\frac{\dis g_A (5 \, + \,4\, g_A \,-\, g_A^2)}
{\dis 12 \, f_A \, f_V} \,, &  Z^{(2)} \, =
\, - \frac{\dis N_c}{\dis 16 \pi^2} \,
\frac{\dis g_A (1 \, + \, g_A)^2}
{\dis 12 \, f_A \, f_V} \,. \hfill \cr\cr } \ee

\vspace*{0.5cm}

\noindent We therefore
 find that these couplings, in the ENJL model
we are considering, are completely fixed by the constants
$f_V$, $f_A$ and $g_A$. This is
due to the fact that the terms in the abnormal
intrinsic parity sector for the interaction of spin-1 fields
with pseudoscalar mesons and external sources
to chiral ${\cal O} (p^3)$ and the WZ effective action
come from the same formal expression.

{}From the expressions in Eq. (\ref{coup}), we find
relations between the various couplings
 that are independent of $g_A$,

\be \label{relab} \matrix{
\frac{\dis h_V}{\dis \sigma_V}
\,=\, \frac{\dis f_V}{\dis \sqrt 2} \hfill \cr\cr
\frac{\dis h_A}{\dis \sigma_A} \,=\, \frac{\dis f_A}{\dis
\sqrt 2} \,. \hfill \cr\cr } \ee

\vspace*{0.5cm}

\noindent These relations appear because of the formal
symmetry between the expression for the
mean-field approximation effective action
and the effective action including spin-1 particle fields
explained in Section \ref{2}. Because of that
symmetry we have, for instance, that
in the effective action $f^{(+)}_{\mu \nu}$
and $V_{\mu \nu}$ always appear in the following
combination

\be \label{combi}
f^{(+)}_{\mu \nu} \,+\, \frac{\dis \sqrt 2}
{\dis f_V} \, V_{\mu \nu} \, ,
\ee

\vspace*{0.5cm}

\noindent therefore the couplings $\sigma_V$
and $h_V$ are related as given in Eq. (\ref{relab}).

The abnormal intrinsic parity couplings involving
spin-1 particles
have been considered in the literature in the context
of different models \cite{ban}-\cite{meis}.
A common feature of these models which differentiates
them from the ENJL cut-off model we are studying
here is that the spin-1 particles are introduced
there as the gauge bosons either of a hidden local symmetry
\cite{ban,fuji} or of the U(3)$_L$ $\times$ U(3)$_R$ chiral
symmetry \cite{kay}. The derivation of the models in Refs.
\cite{ban}-\cite{kay} from an ENJL model in the
presence of the chiral anomaly is studied in
Ref. \cite{waka}. In this Reference,
 special attention is given to the
equivalence of these models for the physics in the
anomalous and non-anomalous sectors of the theory.
The most interesting of these models is the so-called
Hidden Gauge Symmetry (HGS) model \cite{ban,fuji}.
In this model,
vector mesons are the dynamical gauge bosons of
a hidden local U(3)$_V$ symmetry of the chiral Lagrangian.
Ref. \cite{fuji} gives the general solution for
the chiral anomaly  in the
presence of vector mesons when considered as gauge bosons.
The solution is a linear combination of six
invariants which, in general, introduce interaction terms
between Goldstone bosons and external sources apart
{}from those contained in the WZ action. It was also shown
there that these new interaction terms do not change
the low-energy theorems for $\pi^0 \to \gamma \gamma$
and $\gamma \to \pi \pi \pi$ transitions. The ENJL model
we consider here does not introduce any coupling between
Goldstone bosons and external sources apart
{}from those in the WZ action. Thus, in order to
compare both models, we shall require that the linear
combination of six invariants which is a
solution of the anomaly in the HGS model does not introduce
couplings between Goldstone bosons and external sources
apart from those in the WZ action.
In this limit we have that the following relation,
already found in the
ENJL model in Eq. (\ref{relab}),

\be \matrix{
\frac{\dis h_V}{\dis \sigma_V}\,=\,\frac{\dis f_V}{\dis
\sqrt 2}
\hfill\cr\cr}\ee

\noindent is also present in the HGS model. In the same limit
one also finds the following relation in the HGS model

\be \matrix{
\frac{\dis \theta_V}{\dis h_V}\,=\,2\,,
\hfill\cr\cr}\ee

\noindent which in the ENJL model is only true for a
particular value of the $g_A$ coupling. This value
is a solution of the following equation,

\be \label{easo} \matrix{
5\,-\, 7\, g_A\,-\, g_A^2\,-\,g_A^3\,=\,0\,
\hfill \cr\cr} \ee

\vspace*{0.5cm}

\noindent and is $g_A\simeq 0.63$ . (The other two solutions
are not real.)

\setcounter{equation}{0}
\section{The non-anomalous sector} \label{4} \indent

We shall next study the real part of the effective action, {\it
i.e.} the non-anomalous sector of the theory. This sector
can be computed in the ENJL cut-off theory proposed
in Ref. \cite{bbr}, as was shown there, with the help of the
heat kernel expansion \cite{ball}. We shall use the formal
symmetry stated in Section \ref{2}
as we did in the anomalous sector in Section \ref{3}.
Therefore to obtain the effective action in terms of the
spin-1 fields $W_\mu^{(\pm)}$ we only need to calculate the
effective action in the mean-field approximation limit and
replace the external sources by
the primed external sources in Eq. (\ref{92}).
(For a summary of the main technical
details and notation see Ref. \cite{bbr}.)
As explained in Section \ref{2}, from this effective action we can
get the effective action involving the physical spin-1 particle
fields $V_\mu$ and $A_\mu$ by performing
the diagonalization and scale redefinition in Eqs. (\ref{ga}) and
(\ref{scale}).

Within this approach we calculate the non-anomalous couplings of
spin-1
particles to pseudoscalar mesons and external sources at leading
${\cal O}(N_c)$ for terms that
are linear or quadratic in the spin-1 particle fields and to
chiral ${\cal O} (p^3)$. The results for the couplings
in the Lagrangian in Eq. (\ref{47}) are

\be \label{nona1} \matrix{
\alpha_V \, = \,-\, \frac{\dis N_c}{\dis 16 \pi^2} \,
\frac{\dis \sqrt 2 g_A^2}{\dis 6 f_V} \, \left[ \Gamma(0,x) \, - \,
\Gamma(1,x) \right]\,, \hfill \cr\cr
\beta_V \, = \,-\, \frac{\dis N_c}
{\dis 16 \pi^2} \, \frac{\dis \sqrt 2 g_A}{\dis 12 f_V} \, \left[
3 \, \rho \, \Gamma(0,x) \, + \, \Gamma(1,x) \right] \hfill \cr\cr
\hspace*{2cm} {\rm with} \hspace*{1cm} \rho \equiv \frac{\dis M_Q}
{\dis B_0}\,. \hspace*{2cm} \hfill \cr\cr} \ee

\vspace*{0.5cm}

\noindent  For the couplings in the Lagrangian
in Eq. (\ref{48}) we get

\be \label{nona2} \matrix{
\alpha_A \, = \,-\, \frac{\dis N_c}{\dis 16 \pi^2} \,
\frac{\dis \sqrt 2 g_A^2}
{\dis 6 f_A} \, \left[ \Gamma(0,x) \, - \,2\,
\Gamma(1,x) \right]\, , \hfill\cr\cr
\gamma_A^{(1)} \, = \, \frac{\dis N_c}{\dis 16 \pi^2} \,
\frac{\dis \sqrt 2 g_A^2}{\dis 6 f_A} \, \left[ (1 - g_A^2) \,
\Gamma(0,x) \right. \hfill \cr \cr \left.
\hphantom{\gamma_A^{(1)} \, = \, \frac{\dis N_c}{\dis 16 \pi^2} \,
\frac{\dis \sqrt 2 g_A^2}{\dis 12 f_A} \,}
- \,2\, (1\,-\,2\,g_A^2) \, \Gamma(1,x) \, - \, 2 \, g_A^2 \,
\Gamma(2,x) \right]\,, \hfill \cr\cr
\gamma_A^{(2)} \, = \, - \, \frac{\dis N_c}{\dis 16 \pi^2} \,
\frac{\dis \sqrt 2 g_A^2}{\dis 12 f_A} \,
\left[ (1 - g_A^2) \, \Gamma(0,x) \right. \hfill \cr \cr \left.
\hphantom{\gamma_A^{(2)} \, = \, - \, \frac{\dis N_c}{\dis 16 \pi^2}\,
\frac{\dis \sqrt 2 g_A^2}{\dis 6 f_A} \,} - \,2\, (1\,-\,4\,g_A^2) \,
\Gamma(1,x) \, - \, 4\, g_A^2 \, \Gamma(2,x) \right]\,,
\hfill \cr\cr  \gamma_A^{(3)} \, = \,{\cal O}(1/\sqrt{N_c})
 \hspace*{0.5cm} {\rm and} \hspace*{0.5cm}
\gamma_A^{(4)} \, = \,{\cal O} (1/\sqrt{N_c})
 \,. \hfill \cr\cr } \ee

\vspace*{0.5cm}

\noindent  For the couplings in the Lagrangian
in Eq. (\ref{49}) we get

\be \label{nona3} \matrix{
\delta_V^{(1)} \, = \, - \, \frac{\dis N_c}{\dis 16 \pi^2} \,
\frac{\dis
1}{\dis 12 f_V^2} \, \left[ (2\,-\,3\,g_A^2) \, \Gamma(0,x) \,
+ \, 7 \,
g_A^2 \, \Gamma(1,x) \right]\,, \hfill \cr\cr
-\,4\, \delta_V^{(2)} \, = \,-\,4\,\delta_V^{(3)} \, = \,
-\,\delta_V^{(5)} \, =\hfill\cr\cr
\delta_V^{(6)}\,
= \, \frac{\dis N_c}{\dis 16 \pi^2} \,
\frac{\dis g_A^2}{\dis 3 f_V^2} \, \left[ \Gamma(0,x) \, - \,
\Gamma(1,x) \right]\,, \hfill \cr\cr
\delta_V^{(4)} \, = \,{\cal O}(1/N_c) \,, \hfill \cr\cr
\phi_V \, = \, \frac{\dis N_c}{\dis 16 \pi^2} \,
\frac{\dis 1}{\dis 3 f_V^2} \, \left[ \Gamma(0,x) \, -\, \Gamma(1,x)
\right] \,. \hfill \cr\cr} \ee

\vspace*{0.5cm}

\noindent  For the couplings in the Lagrangian
in Eq. (\ref{50}) we get

\be \label{nona4} \matrix{
\delta_A^{(1)} \, = \, - \, \frac{\dis N_c}{\dis 16 \pi^2} \,
\frac{\dis
g_A^2}{\dis 12 f_A^2} \, \left[ (2\,-\,3\,g_A^2) \, \Gamma(0,x)
\right. \hfill \cr \cr \left. \hphantom{\delta_A^{(1)} \,=\, - \,
\frac{\dis N_c}{\dis 16 \pi^2} \,
\frac{\dis g_A^2}{\dis 12 f_A^2} \,} - \,
4 \, (1\,-\,6\,g_A^2)\, \Gamma(1,x) \,-\, 12\, g_A^2 \, \Gamma(2,x)
\right]\,, \hfill \cr\cr
\delta_A^{(2)} \, = \,
\delta_A^{(3)} \, = \, - \, \frac{\dis N_c}{\dis 16 \pi^2} \,
\frac{\dis g_A^4}{\dis 12 f_A^2} \, \left[ \Gamma(0,x) \,-\, 8 \,
\Gamma(1,x) \,-\, 4 \, \Gamma(2,x) \right]\,, \hfill \cr\cr
\delta_A^{(4)} \, = \,{\cal O}(1/N_c) \,, \hfill \cr\cr
\delta_A^{(5)} \, = \,-\, \frac{\dis N_c}{\dis 16 \pi^2} \,
\frac{\dis g_A^4}
{\dis 3 f_A^2} \, \left[ \Gamma(0,x) \, - \,8\, \Gamma(1,x) \, +\,
4\, \Gamma(2,x) \right]\,, \hfill \cr\cr
\delta_A^{(6)} \, = \, \frac{\dis N_c}{\dis 16 \pi^2} \,
\frac{\dis g_A^4}{\dis 3 f_A^2} \,
\left[ \Gamma(0,x) \, - \,4\, \Gamma(1,x) \,+\,2\, \Gamma(2,x)
\right]\,, \hfill \cr\cr
\phi_A \, =\, \frac{\dis N_c}{\dis 16 \pi^2}
\, \frac{\dis g_A^2}{\dis 3 f_A^2} \, \Gamma(0,x)\,.
 \hfill \cr\cr} \ee

\vspace*{0.5cm}

\noindent  For the couplings in the Lagrangian
in Eq. (\ref{51}) we get

\be \label{nona5} \matrix{
A^{(1)} \, = \,
A^{(2)} \, = \,-\, \frac{\dis N_c}{\dis 16 \pi^2}
\, \frac{\dis g_A^2}{\dis 3 f_A f_V} \, \left[ \Gamma(0,x) \,-\,
\Gamma(1,x) \right]\,, \hfill \cr\cr
A^{(3)} \, = \,-\, \frac{\dis N_c}{\dis 16 \pi^2} \,
\frac{\dis g_A^2}{\dis
3 f_A f_V} \, \left[ \Gamma(0,x) \, - \, 2 \, \Gamma(1,x) \right]\,,
\hfill \cr\cr
B \, = \,-\, \frac{\dis N_c}{\dis 16 \pi^2} \, \frac{\dis g_A}
{\dis 2 f_A f_V} \, \rho \, \Gamma(0,x)\,. \hfill \cr\cr } \ee

\vspace*{0.5cm}

\noindent  Finally, for the couplings in the Lagrangian
in Eq. (\ref{52}) we get

\be \label{nona6} \matrix{
\hspace*{3cm}
C^{(1)}\,=\,- \frac{\dis \sqrt 2}{\dis f_A} \, c_d \,,
\hspace*{2cm}
C^{(2)}\,=\, \frac{\dis 4}{\dis f_A^2} \, c_d\hfill\cr\cr
{\rm with} \hspace*{2cm} c_d\,=\, \frac{\dis
N_c}{\dis 16 \pi^2} \, \frac{\dis M_Q}{\dis \lambda_S}
\, 2 \, g_A^2 \, \left[ \Gamma(0,x)\,-\, \Gamma(1,x)\right] \,,
\hspace*{3cm} \hfill\cr\cr \hspace*{3cm}
D\,=\,0\,. \hfill \cr\cr } \ee

\vspace*{0.5cm}

\noindent From the expressions
in Eqs. (\ref{nona1})-(\ref{nona6})
we can obtain the following relations that
are independent of $g_A$,

\be \label{relab2} \matrix{
\frac{\dis \alpha_V}{\dis A^{(1)}} \,=\, \frac{\dis \alpha_V}
{\dis A^{(2)}} \,=\, \frac{\dis f_A}{\dis \sqrt 2} \hfill \cr\cr
\frac{\dis \alpha_A}{\dis A^{(3)}} \,=\, \frac{\dis f_V}{\dis
\sqrt 2}  \hfill \cr\cr \frac{\dis C^{(1)}}{\dis C^{(2)}} \,=\,
-\,\frac{\dis f_A}{\dis 2\,\sqrt 2}  \hfill \cr\cr} \ee

\noindent which are also due to the formal symmetry explained
in Section \ref{2}, like the relations in Eq. (\ref{relab}).

\setcounter{equation}{0}
\section{Phenomenological applications} \label{5} \indent

In this Section we shall discuss some phenomenological
applications from our calculations. In
the whole following analysis we shall work in the chiral limit,
{\it i.e.} ${\cal M} \to 0$. Therefore, we shall disregard
possible $\omega_8-\rho^0$ and $\eta_8-\pi^0$
mixings, which are proportional to light-quark masses. In
addition, the operators that generate these mixings
are order ${\cal O} (p^4)$ in the chiral
expansion \cite{gl2}. Here we shall  use the
coupling constants we have calculated in the previous Sections
in the ENJL cut-off model \cite{bbr} to make predictions
on a large variety of processes where spin-1 particle
are involved (anomalous transitions,
radiative decays, $\cdots$).
The radiative decays of vector mesons in the context
of Zweig's rule and explicit SU(3)-symmetry violation
induced by a non-vanishing strange-quark mass
was considered in Ref. \cite{chiv} in
the non-relativistic quark model. The different
coupling constants appearing there
were fixed from a fit to experimental data. Phenomenology involving
spin-1 particles has been discussed before within different
approaches. See for instance  Refs. \cite{chiv}-
\cite{ivan}.

\subsection{Vector resonance decays} \label{6.1} \indent

Here we shall discuss the predictions for
vector particle decays. As was already mentioned
in Section \ref{2}, in the ENJL model we are considering,
all the couplings can be written in
terms of three parameters, namely $g_A$, $M_Q$ and
$\Lambda_\chi$. The coupling $g_A$ was defined in Eq. (\ref{ga})
and in this model takes the following value \cite{bbr}

\be \label{gad}
g_A\,=\,1\,-\,\frac{\dis f_\pi^2}{\dis f_V^2 \, M_V^2}\, .
\ee

\vspace*{0.5cm}

\noindent We shall choose for $g_A$, $M_Q$ and $\Lambda_\chi$
 the values that are obtained from
fitting the ${\cal O} (p^2)$ and
${\cal O} (p^4)$ couplings of the chiral Lagrangian
\cite{bbr}, {\it i.e.}

\be \label{param} \matrix{
g_A \,=\,0.65\,, \hspace*{0.5cm} x\equiv \frac{\dis M_Q^2}
{\dis \Lambda_\chi^2} \,=\,0.06 \hspace*{0.5cm}
{\rm and} \hspace*{0.5cm} M_Q\,=\,260 \,\,{\rm MeV} \,.} \ee

\vspace*{0.5cm}

\noindent These values correspond to $f_V \,=\, 0.17$,
$g_V \,=\, 0.083$ and $f_A \,= \,0.080$.

For the vector mixing $\omega - \phi$ we use the ideal
angle, {\it i.e.} $\tan \varphi_V \,=\, 1/\sqrt 2$ and for
the pseudoscalar mixing we use $\tan \varphi_P \,=\, -\,
1/2 \sqrt 2$. In both cases we assume nonet symmetry.
The diagonalized $\omega$, $\phi$, $\eta$ and $\eta'$
states in terms of the SU(3) octet and singlet states in
Eqs. (\ref{eta8}), (\ref{eta1}), (\ref{w8}) and (\ref{w1})
are

\be \label{diag} \matrix{
\eta' \,=\, \cos \varphi_P \,\, \eta_1 \,+\, \sin \varphi_P
\,\, \eta_8 \,, \hfill \cr
\eta \,=\, -\, \sin \varphi_P \,\, \eta_1 \,+\, \cos \varphi_P
\,\, \eta_8 \,, \hfill \cr\cr
\omega \,=\, \cos \varphi_V \,\, \omega_1 \,+\, \sin \varphi_V
\,\, \omega_8 \,, \hfill \cr
\phi \,=\, -\, \sin \varphi_V \,\, \omega_1 \,+\, \cos \varphi_V
\,\, \omega_8 \,. \hfill \cr\cr } \ee

First we shall study the decays $V \to P \gamma$
($P \to V \gamma$), for which the Feynman diagrams
are shown in Figure \ref{fig1}.
The amplitude for the $V \to P \gamma$
decay is given by the expression

\be \label{vpga} \matrix{
A(V \to P (p) \gamma (k)) \,=\,
C_{VP\gamma} \,4 \, |e| \, \sqrt 2 \,
\frac{\dis h_V}{\dis f_\pi} \, \epsilon_{\mu \nu \alpha
\beta} \, \varepsilon^{\mu}_{(\gamma)} \, k^\nu \,
\varepsilon^{*\alpha}_{(V)} \, p^\beta \hfill \cr\cr
\hphantom{A(V \to P (p) \gamma (k)) \,=\,
C_{VP\gamma} }\,\times \, \left[ 1 \,+ \, \sqrt 2
\, \frac{\dis f_V \sigma_V}
{\dis h_V} \, \frac{\dis k^2}{\dis M_{V'}^2 \,-\,
k^2 \,-\, i \, M_{V'} \Gamma_{V'}} \right] \,.
\hfill \cr \cr} \ee

\noindent Here, $C_{VP\gamma}$ is an SU(3)-light-flavour
symmetry factor that relates the different amplitudes.
 The different values for this factor and the
intermediate resonance $V'$ appearing in Eq. (\ref{vpga})
for each process are given in Table \ref{table1b}.
We denote the polarization pseudovector
corresponding to the particle $V$ by $\varepsilon^\mu_
{(V)}$. The decay rate corresponding
to the amplitude in Eq. (\ref{vpga}) is

\be \label{vpgr} \matrix{
\Gamma(V \to P \gamma) \,=\, |C_{VP\gamma}|^2 \,
\frac{\dis 4 \, \alpha \, h_V^2}{\dis 3} \,
\frac{\dis M_V^3}{\dis f_\pi^2} \,
\lambda^{3/2} \left(1, \frac{\dis m_P^2}{\dis
M_V^2}, \frac{\dis k^2}{\dis M_V^2} \right) \hfill \cr\cr
\hphantom{\Gamma(V \to P \gamma) \,=\, |C_{VP\gamma}|^2 \,
\frac{\dis 4 \alpha h_V^2}{\dis 3}} \times\,
 \left| 1 \,+ \, \frac{\dis 2 \, k^2}{\dis M_{V'}^2
\,-\, k^2 \,-\, i \, M_{V'} \Gamma_{V'}} \right|^2 \, ,
\hfill \cr\cr }\ee

\noindent
with $\alpha \,= \, \frac{\dis e^2}{\dis 4 \pi}$
and

\be
\lambda(x,y,z) \,=\, x^2 \,+\, y^2 \,+\, z^2 \,-\,2\,
x y \,-\,2\, x z \,-\,2\, y z \,.
\ee

\vspace*{0.5cm}

\begin{table}[htb]
\caption{SU(3)-symmetry factors ($C_{VP\gamma}$)
and intermediate resonance ($V'$)
for the $V \to P \gamma$ ($P\to V \gamma$) decays.}

\label{table1b}
\begin{center}
\begin{tabular}{|l|c|c|}
\hline
 Process & C$_{VP\gamma}$ & $V'$\\
\hline
\hline
$\rho \to \pi \gamma$ &1/3&
$\omega_8$ \\
$\omega \to \pi^0 \gamma$ & 1 & $\rho^0$\\
$\rho^0 \to \eta \gamma$ & $-\, \sqrt 2 / \sqrt 3$
& $\rho^0$\\
$\omega \to \eta \gamma$ & $\sqrt 2 / (3 \sqrt 3)$
& $\omega_8$\\
$\eta' \to \rho^0 \gamma$ & 1 & $\rho^0$\\
$\eta' \to \omega \gamma$ &$1 / 3$ & $\omega_8$\\
$\phi \to \pi^0 \gamma$ & 0 & $\rho^0$\\
$\phi \to \eta \gamma$ & $2/ (3 \sqrt 3)$
& $\omega_8$\\
$\phi \to \eta' \gamma$&$-\, 2 \sqrt 2 /
\sqrt 3$ & $\omega_8$\\
$K^{* 0} \to K^0 \gamma$ & $2 / 3$& ---\\
$K^{* +} \to K^+ \gamma$ & $1 / 3$& ---\\
\hline
\end{tabular} \end{center} \vspace*{0.5cm}\end{table}

The decay rates for the $\omega$ and $\rho^0$ vectors
can have contributions from a possible $\omega_8-\rho^0$ mixing.
In addition, the amplitude for
$\rho \to \pi \gamma$ has a sizable contribution from the
absortive part of pion and kaon chiral loops (see Figure \ref{fig2}).
(For the other processes this contribution is negligible.)
We have taken into account this last contribution which
has the following expression

\vspace*{0.5cm}

\be \label{abs}
\Gamma(\rho \to \pi \gamma){\dis |}_{\rm Abs}
 \,=\,
\frac{\dis \alpha \,g_V^2\,M_\rho^{11}}{\dis (768 \sqrt 6)^2
\, \pi^6 \, f_\pi^{10}} \, \left(1\,-\, \frac{\dis
m_\pi^2}{\dis M_\rho^2} \right)^3 \, \left(1\,-\, \frac{\dis
4 m_\pi^2}{\dis M_\rho^2}\right)^3 \,.
\ee

\vspace*{0.5cm}

\noindent The predictions listed in the
column Prediction 1 in Table
\ref{table1} correspond to the values of $g_A$, $M_Q$ and $x$
in Eq. (\ref{param}). The worst results in this column,
taking into account the experimental errors, are for the decays
$\phi \to \eta \gamma$ and $K^{* +} \to K^+ \gamma$ which are
between 4 and 6 standard deviations ($\sigma$)  from the experimental
value. These disintegrations, however, can be explained with the
inclusion of explicit chiral symmetry breaking terms
proportional to the strange-quark mass as shown in the
non-relativistic quark model in Ref. \cite{chiv}.
The predictions for the other processes are less than 2 $\sigma$
{}from the central experimental values.  The processes $\rho^0
\to \pi^0 \gamma$ and $\omega \to \pi^0 \gamma$ are likely to be
influenced by a possible $\omega_8-\rho^0$ mixing. All the predictions
for the transitions in Table \ref{table1} are made in the chiral limit
and depend on two couplings, $h_V$ and $g_V$; in fact the constant
$g_V$ only affects to $\rho \to \pi \gamma$ decays and its
contribution is not dominant. We  refrain from
doing a fit of the experimental results since all the
amplitudes are related by SU(3)-symmetry factors (up to
the small part proportional to $g_V$ in $\rho \to \pi \gamma$
decays) and higher order chiral corrections
for these processes are very different for each of them.

\begin{table}[htb]
\caption{Partial widths in keV corresponding to $V \to P \gamma$
and $P \to V \gamma$ decays.}

\label{table1}
\begin{center}
\begin{tabular}{|l|c|c|c|}
\hline
 Process & Prediction 1& Prediction 2& Experiment\\
\hline
\hline
$\rho^+ \to \pi^+ \gamma$ & 58 & (${}^\dagger$)
&68 $\pm$ 7 \\
$\rho^0 \to \pi^0 \gamma$ & 58 & 68 &120 $\pm$ 30 \\
$\omega \to \pi^0 \gamma$ & 452 & 546 &717 $\pm$ 50 \\
$\rho^0 \to \eta \gamma$ & 35 & 42 &58 $\pm$ 11 \\
$\omega \to \eta \gamma$ & 3 & 4 & 4 $\pm$ 2 \\
$\phi \to \eta \gamma$ & 75(${}^*$)& 91(${}^*$) &57 $\pm$ 3 \\
$\eta' \to \rho^0 \gamma$ & 41 & 50& 59 $\pm$ 9 \\
$\eta' \to \omega \gamma$ & 4 & 5&6 $\pm$ 1 \\
$K^{* 0} \to K^0 \gamma$ & 110(${}^*$)& 133(${}^*$)&116 $\pm$ 12 \\
$K^{* +} \to K^+ \gamma$ & 28(${}^*$)& 34(${}^*$)&50 $\pm$ 6 \\
\hline
\end{tabular}
\end{center}

(${}^\dagger$) Input to fix $g_A$ ($g_A\,=$ 0.76).

(${}^*$) These processes get contributions
{}from explicit chiral symmetry breaking terms
proportional to the strange-quark mass \cite{chiv}.
\vspace*{0.5cm} \end{table}

Instead, the most reliable prediction is
for the $\rho^+ \to \pi^+ \gamma$ decay because is not
affected by neutral particle mixings and the
explicit chiral symmetry breaking is small.
All the other transitions are either affected by mixings or by
sizable explicit chiral symmetry breaking effects. The
agreement between the result for the $\rho^+ \to \pi^+ \gamma$
decay in the column Prediction 1 and experiment can be improved by
using either a larger value for $g_A$ or a smaller value for
 $f_V$ ({\it i.e.}
a larger value of $x$) or both. Given the poor knowledge of
the parameter $g_A$ and the rather good value for $f_V$
we are using ($f_V\,=$ 0.17 compared to the experimental value
$f_V {\dis |}_{\rm exp} \simeq 0.20$),
we shall fix the value of $f_V$ to $f_V\,=$ 0.17
and fix from the decay
$\rho^+ \to \pi^+ \gamma$ the value of $g_A$.
We find then $g_A \,=\, 0.76$. So that,
this value of $g_A$ leads to
$h_V\,=\,0.033$. We shall use this value to see how
the predictions in Table \ref{table1} vary with $g_A$.
The corresponding results with $g_A\,=\,0.76$ are in the column
Prediction 2. We find that, in general,
the results are improved, except for the decay
 $\phi \to \eta \gamma$
which is now 10 $\sigma$ from the central experimental value.
However, as was already stated, this decay rate is very
sensitive to explicit chiral symmetry breaking.
The prediction for the decay $\omega
\to \pi^0 \gamma$ improves but is still 3.5 $\sigma$ from
the experimental value which probably means that the
$\omega_8-\rho^0$ mixing is likely to be important.

Next, we shall study the predictions for the $V \to \pi \pi \pi$
decays for which the results
can be found in Table \ref{table2}. The Feynman
diagrams for this process are shown in Figure \ref{fig3}.
The amplitude for $\omega \to \pi^+ \pi^- \pi^0$
has the form,

\be \label{w3p} \matrix{
A(\omega \to \pi^+(p_1) \pi^- (p_2) \pi^0(p_3)) \,=\,
\frac{\dis 6 \sqrt 2}{\dis f_\pi^3} \, \theta_V \,
\epsilon_{\mu \nu \alpha \beta} \,
\varepsilon_{(\omega)}^{*\mu} \, p_1^\nu \, p_2^\alpha \,
p_3^\beta \, \hfill \cr\cr \hphantom{A(\omega \to
\pi^+(p_1) \pi^- (p_2) } \times \,
\left[1\,+\,
\frac{\dis 4 \, \sqrt 2}{\dis 3} \, \frac{\dis g_V \sigma_V}
{\dis \theta_V} \, {\dis \sum^{i<j}_{i,j=1,2,3}} \,
\frac{\dis p_{ij}^2}
{\dis M_\rho^2 \,-\, p_{ij}^2 \, -\,i \, M_\rho \Gamma_\rho}
\right]
\hfill \cr\cr } \ee

\vspace*{0.5cm}

\noindent with $p_{ij}\,=\,p_i\,+\,p_j$.
The amplitudes for $\rho \to \pi \pi \pi$ decays vanish
at lowest order. The decay rate for $\omega \to \pi^+ \pi^-
\pi^0$ is

\vspace*{0.5cm}

\be \label{inte}
\Gamma(\omega \to \pi^+ \pi^- \pi^0) \, = \,
\frac{\dis 1}{\dis 384 \, \pi^3 M_\omega} \, {\dis \int}
\, {\rm d} E_+ \, {\dis \int} \, {\rm d} E_- \,
{\dis \sum_{\rm pol}} \, \left|A \right|^2
\ee

\vspace*{0.5cm}

\noindent with $E_\pm \equiv E_{\pi^+} \pm E_{\pi^-}$.
The limits in Eq. (\ref{inte}) are given by

\be \label{limitsa} \matrix{
(E_+)_{\rm max} \, = \, M_\omega \, - \, m_\pi \,,
\hspace*{1.5cm}
(E_+)_{\rm min} \,=\, \frac{\dis M_\omega}{\dis 2} \,+\,
\frac{\dis 3 m_\pi^2}{\dis 2 M_\omega} \,,\hfill \cr\cr
(E_-)_{{\rm max}}
\,=\, \frac{\dis m_\pi^2 + M_\omega^2 - 2 x - 2
 \left(M_{\pi^0\pi^-}^2\right)_{{\rm min}}}
{\dis 2 M_\omega}\,, \hfill \cr\cr
(E_-)_{{\rm min}}
\,=\, \frac{\dis m_\pi^2 + M_\omega^2 - 2 x - 2
 \left(M_{\pi^0\pi^-}^2\right)_{{\rm max}}}
{\dis 2 M_\omega}\,, \hfill \cr\cr} \ee

\noindent with

\be \label{limitsb} \matrix{
\left(M_{\pi^0\pi^-}^2\right)_{{\rm min}({\rm max})}  \,= \,
\frac{\dis 1}{\dis 8 (m_\pi^2 + x)} \, \left[
\left(M_\omega^2 - m_\pi^2\right)^2 \right. \hfill \cr\cr
\hspace*{0.1cm} \left. \,-\,
\left( \lambda^{1/2}(2(m_\pi^2+x), m_\pi^2, m_\pi^2)
+(-) \, \lambda^{1/2}(M_\omega^2, m_\pi^2, 2 (m_\pi^2+x))
\right)\right] \,, \hfill \cr\cr
x\equiv p_1 \cdot p_2
\,=\, M_\omega E_+ - \frac{\dis M_\omega^2 + m_\pi^2}
{\dis 2}\,. \hfill \cr\cr} \ee

\begin{table}[htb]
\caption{Partial widths in keV corresponding to $V \to \pi \pi \pi$,
$V \to P_1 P_2 \gamma$, $V \to V' \gamma$ and $V \to V' \pi$
decays.}

\vspace*{0.3cm}

\label{table2}
\begin{center}
\begin{tabular}{|l|c|c|c|}
\hline
 Process & Prediction 1 & Prediction 2 &Experiment\\
\hline
\hline
$\omega \to \pi^+ \pi^- \pi^0$ & 6.8 $\cdot$ 10$^3$ &
7.5 $\cdot$ 10$^3$&
(7.5 $\pm$ 0.1) $\cdot$ 10$^3$\\
$\rho^0 \to \pi^+ \pi^- \pi^0$ & 0 & 0&$<$ 20 \\
$\rho^+ \to \pi^+ \pi^- \pi^+$ & 0 & 0&--- \\
\hline
$\rho^+ \to \pi^+ \pi^0 \gamma$ & 31 &29& --- \\
$\rho^0 \to \pi^+ \pi^- \gamma$ & 4 & 7.5& (${}^\#)$ \\
$\rho^0 \to \pi^0 \pi^0 \gamma$ & 0.5 &0.7& --- \\
$\rho^0 \to \pi^0 \eta \gamma$ & 2 $\cdot$ 10$^{-5}$ &
3 $\cdot$ 10$^{-5}$ &--- \\
$\omega \to \pi^+ \pi^- \gamma$ & 0.16 & 0.23 & $<$ 30 \\
$\omega \to \pi^0 \pi^0 \gamma$ & 0.08 & 0.12 &$<$ 3.4 \\
$\omega \to \pi^0 \eta \gamma$ & 5 $\cdot$ 10$^{-4}$ &
7 $\cdot$ 10$^{-4}$& --- \\
$\phi \to \overline K^0 K^0 \gamma$ & 4 $\cdot$ 10$^{-9}$
& 6 $\cdot$ 10$^{-9}$ & --- \\ \hline
$\omega \to \rho^0 \gamma$ &0&0& --- \\
$\phi \to \omega \gamma$ &0 &0& $<$ 220 \\
$\phi \to \rho^0 \gamma$ &0 &0& $<$ 90 \\
$\phi \to \omega \pi^0$ & 0 &0& --- \\
\hline
\end{tabular}
\end{center}
(${}^\#$) This transition is dominated by bremsstrahlung off
pions. In Ref. \cite{dol} an upper bound of 0.76 MeV is given
for the structural bremsstrahlung.
\vspace*{0.5cm} \end{table}

\vspace*{0.5cm}

\noindent The results in the column Prediction 1 in Table
\ref{table2} are for the input values of $g_A$, $M_Q$ and
$x$ in Eq. (\ref{param}). Here,  the experimental knowledge
is very limited and there exist just upper bounds for several
of the transitions. Only the decay
rate for $\omega \to \pi^+ \pi^- \pi^0$ is known.
In our model, this amplitude depends on the
coupling constants $\theta_V$ and $\sigma_V$.
With the value of $g_A$ from $\rho^+ \to \pi^+ \gamma$
($g_A\,=\,0.76$), our predictions are those in
the column Prediction 2 in Table \ref{table2}.

We next proceed to the study of the $V \to P_1 P_2 \gamma$ decays. The
predictions we get for them are also shown in Table \ref{table2}
and the corresponding Feynman diagrams shown
in Figure \ref{fig4}.
We shall study first the decays with charged pseudoscalar
mesons in the final state.
The amplitude we obtain for the $\rho^0 \to \pi^+ \pi^- \gamma$
decay is

\be \label{rppg0} \matrix{
A(\rho^0 \to \pi^+(p_1) \pi^-(p_2) \gamma(k)) \,=\,
2 \sqrt 2 \, |e| \, \frac{\dis \alpha_V}{\dis f_\pi^2} \,
\varepsilon^{* \mu}_{(\rho)} \,
\left(\varepsilon_\mu^{(\gamma)}
 \, k_\nu - \varepsilon_\nu^{(\gamma)} \, k_\mu\right) \,
\left(p_1+p_2\right)^\nu \,; \hfill \cr\cr } \ee

\vspace*{0.5cm}

\noindent for the $\rho^+ \to \pi^+ \pi^0 \gamma$ decay,

\be \label{rppg+} \matrix{
A(\rho^+ \to \pi^+(p_1) \pi^0(p_2) \gamma(k)) \,=\,
2 \sqrt 2 \, |e| \, \frac{\dis \alpha_V}{\dis f_\pi^2} \,
\varepsilon^{* \mu}_{(\rho)} \, \left(
\varepsilon_\mu^{(\gamma)} \, k_\nu -
\varepsilon_\nu^{(\gamma)}
 \, k_\mu \right) \hfill\cr\cr \hspace*{1cm} \times \,
\left[ p_2^\nu \,
+\, \sqrt 2 \frac{\dis g_V \, \phi_V}{\dis \alpha_V} \,
\frac{\dis (p_1+p_2)^2 \,(p_2-p_1)^\nu}{\dis M_\rho^2 -
(p_1+p_2)^2 - i M_\rho \Gamma_\rho}\right] \hfill\cr\cr}
\ee

\vspace*{0.5cm}

\noindent and for the $\omega \to \pi^+ \pi^- \gamma$
decay,

\be \label{w2pg+} \matrix{
A(\omega \to \pi^+(p_1) \pi^-(p_2) \gamma(k)) \,=\,
- \, \frac{\dis 16 \sqrt 2}{\dis 3} \, |e| \, \frac{\dis
\sigma_V \, h_V}{\dis f_\pi^2} \, \epsilon_{\mu\nu
\alpha\beta} \, \epsilon^\beta_{. \, bcd}
\,\varepsilon_{(\omega)}
^{*\mu} \, \varepsilon_{(\gamma)}^c \, k^d \, \hfill\cr\cr
\hspace*{1.5cm} \times \, \left(
p_1^\nu \, (p_2+k)^\alpha \, p_2^b \, \frac{\dis 1}{\dis
M_\rho^2 - (p_2+k)^2 - i M_\rho \Gamma_\rho} \,
+ \, (p_1 \leftrightarrow p_2) \right)\,.
\hfill \cr\cr } \ee

\noindent The amplitudes corresponding
to the decays $V^0 \to P^0 P^0 \gamma$ are

\be \label{vppg0} \matrix{
A(V^0 \to P^0(p_1) P^0(p_2) \gamma(k)) \,=\,
- \, C_{VPP\gamma} \,
16 \, \sqrt 2\, |e| \, \frac{\dis
\sigma_V \, h_V}{\dis f_\pi^2} \, \epsilon_{\mu\nu
\alpha\beta} \, \epsilon^\beta_{.\, bcd}
\,\varepsilon_{(V)}
^{*\mu} \, \varepsilon_{(\gamma)}^c \, k^d \, \hfill\cr\cr
\hspace*{0.5cm} \times \, \left(
p_1^\nu \, (p_2+k)^\alpha \, p_2^b \, \frac{\dis 1}{\dis
M_{V'}^2 - (p_2+k)^2 - i M_{V'} \Gamma_{V'}} \,
+ \, \left(\begin{array}{ccc} p_1 & \leftrightarrow & p_2\\
V'& \rightarrow & V''\ea \right) \right)\,.
\hfill \cr\cr }\ee

\vspace*{0.5cm}

\noindent The SU(3)-symmetry factors $C_{VPP\gamma}$
and the intermediate
resonances $V'$ and $V''$ are listed in Table \ref{table2b}.

\begin{table}[htb]
\caption{SU(3)-symmetry factors ($C_{VPP\gamma}$)
and intermediate resonances ($V'$, $V''$)
for the $V^0 \to P^0 P^0 \gamma$ decays.}

\label{table2b}
\begin{center}
\begin{tabular}{|l|c|c|}
\hline
 Process & C$_{VPP\gamma}$ & $V'$, $V''$\\
\hline
\hline
$\rho^0 \to \pi^0 \pi^0 \gamma$ &1&
$\omega, \omega$ \\
$\omega \to \pi^0 \pi^0 \gamma$ &
$1 / 3$ & $\rho^0,\rho^0$\\
$\rho^0 \to \pi^0 \eta \gamma$ & $\sqrt 2 / (3 \sqrt 3)$
 & $\rho^0, \omega$\\
$\omega \to \pi^0 \eta \gamma$ & $\sqrt 2 / \sqrt 3$
 & $\omega, \rho^0$\\
$\phi \to \pi^0 \pi^0 \gamma$& 0 & $\rho^0, \rho^0$\\
$\phi \to \pi^0 \eta \gamma$ & 0
& $\omega, \rho^0$\\
$\phi \to K^0 \overline{K^0} \gamma$&$-\, \sqrt 2 / 3$
 & $K^{* 0}, \overline{K^{* 0}}$\\
\hline
\end{tabular}
\end{center} \vspace*{0.5cm} \end{table}

\vspace*{0.5cm}

The decay rate for all the $V \to P_1 P_2 \gamma$
decays can be written as follows

\be \label{vppg}
\Gamma(V \to P_1 P_2 \gamma) \, = \,
\frac{\dis 1}{\dis 384 \, \pi^3 M_V} \, {\dis \int}
\, {\rm d} E_+ \, {\dis \int} \, {\rm d} E_- \,
{\dis \sum_{\rm pol}} \, \left|A \right|^2
\ee

\vspace*{0.5cm}

\noindent with $E_\pm \equiv E_{P_1} \pm E_{P_2}$.
The limits in Eq. (\ref{vppg}) are given by

\be \label{limits2a} \matrix{
(E_+)_{\rm max} \, = \, M_V \,,\hspace*{1.5cm}
(E_+)_{\rm min} \,=\, \frac{\dis M_V}{\dis 2} \,+\,
\frac{\dis (m_{P_1}+m_{P_2})^2}{\dis 2 M_V}
\,,\hfill \cr\cr
(E_-)_{{\rm max}}
\,=\, \frac{\dis m_{P_1}^2 - m_{P_2}^2 +
M_V^2 - 2 x - 2
 \left(M_{\gamma P_2}^2\right)_{{\rm min}}}
{\dis 2 M_V \hphantom{min}}\,, \hfill \cr\cr
(E_-)_{{\rm min}}
\,=\, \frac{\dis m_{P_1}^2 - m_{P_2}^2 +
M_V^2 - 2 x - 2
 \left(M_{\gamma P_2}^2\right)_{{\rm max}}}
{\dis 2 M_V \hphantom{max}}\,, \hfill \cr\cr } \ee

\noindent with

\be \label{limits2b} \matrix{
\left(M_{\gamma P_2}^2\right)_{{\rm min}({\rm max})}  \,= \,
\frac{\dis 1}{\dis 4(m_{P_1}^2+m_{P_2}^2+2 x)} \, \left[
(M_V^2+m_{P_2}^2-m_{P_1}^2)^2 \right. \hfill \cr\cr
\left. \hspace*{0.1cm} \,-\,
\left( \lambda^{1/2}(m_{P_1}^2+m_{P_2}^2+2x, m_{P_1}^2,
m_{P_2}^2)
+(-) \, \lambda^{1/2}(M_V^2, 0, m_{P_1}^2+m_{P_2}^2+2x)
\right)\right] \,, \hfill \cr\cr
x\equiv p_1 \cdot p_2\,=\, M_V E_+ -
\frac{\dis M_V^2 + m_{P_1}^2 + m_{P_2}^2}
{\dis 2}\,. \hfill \cr\cr} \ee

\vspace*{0.5cm}

\noindent For the transitions with two identical particles
in the final state there is an additional factor 1/2 in the
decay rate in Eq. (\ref{vppg}).
The amplitudes for the $\omega \to \rho^0
\gamma$, $\phi \to \omega \gamma$, $\phi \to
\rho^0 \gamma$ and $\phi \to \omega \pi^0$ decays are
 zero at lowest order.

Up to now we have considered the $\omega-\phi$ mixing
angle to be ideal. The $\phi \to \pi^0
\gamma$, $\phi \to \rho^+ \pi^-$,
$\phi \to \pi^+ \pi^- \pi^0$
and $\phi \to P^0 P^0 \gamma$ decays are proportional
to the deviation of the actual $\omega-\phi$ mixing angle from
the ideal one. The effect of a small deviation on
the other $\phi$ vector decays studied before is not
sizable. We define the deviation of the
$\omega - \phi$ mixing angle from
the ideal one as the difference

\be
\epsilon \equiv \varphi_V \,-\, \varphi \,,\ee

\noindent where $\varphi$ is the physical $\omega - \phi$
mixing angle and $\varphi_V$ is the ideal mixing angle.
The amplitude for $\phi \to \rho^+ \pi^-$ we obtain is

\be \label{firhop} \matrix{
A(\phi \to \rho^+(p_1) \pi^-(p_2)) \,=\, \epsilon \, \,
\frac{\dis 8}{\dis f_\pi} \, \sigma_V\, \epsilon_{\mu
\nu \alpha \beta}\, \varepsilon^{*\mu}_{(\phi)} \, p_1^\nu
\, \varepsilon^\alpha_{(\rho)}\, p_2^\beta\,,
\hfill \cr\cr }\ee

\vspace*{0.5cm}

\noindent giving the following  decay rate

\be \label{firate} \matrix{
\Gamma \,=\, \epsilon^2 \,\, \frac{\dis 2}{\dis 3 \pi}
\, \sigma_V^2 \, \frac{\dis M_\phi^3}{\dis f_\pi^2} \,
\lambda^{3/2} \left(1, \frac{\dis M_\rho^2}{\dis M_\phi^2},
\frac{\dis m_\pi^2}{\dis M_\phi^2}\right) \, .
\hfill\cr\cr}\ee

\vspace*{0.5cm}

\noindent The amplitudes for the other
transitions which are proportional to this deviation
can be easily obtained by multiplying
the corresponding transition amplitudes of the
$\omega$ vector by a factor $\epsilon$.
 The decay rates are the corresponding ones
obtained by changing the $\omega$ vector parameters
by the $\phi$ vector ones. We can predict this deviation
if we assume that the measured partial
widths in Ref. \cite{pdg}
for these transitions come only from a non-ideal mixing.
First, we shall use the set of parameteres in Eq. (\ref{param})
for the input values of $g_A$, $M_Q$ and $x$.
{}From $\phi \to \pi^0 \gamma$ we then obtain
the following result

\be
\epsilon^2 \,=\, (5.7 \pm 0.8) \cdot 10^{-3} \,,\ee

\noindent and from $\phi \to \rho^+ \pi^-$

\be
\epsilon^2 \,=\, (7.6 \pm 0.5) \cdot 10^{-3} \,.\ee

\noindent Using the average of these two results
we predict the decay rates for the $\phi$ vector that
are in the column Prediction 1 in Table \ref{table4}.

\begin{table}[htb]
\caption{Partial widths in keV corresponding to $\phi$-decays
assuming non-ideal $\omega-\phi$ mixing.}

\vspace*{0.3cm}

\label{table4}
\begin{center}
\begin{tabular}{|l|c|c|c|}
\hline
 Process & Prediction 1 & Prediction 2 & Experiment\\
\hline
\hline
$\phi \to \pi^0 \gamma$ & (${}^\dagger$) & (${}^\dagger$) &
5.8 $\pm$ 0.8 \\
$\phi \to \rho^+ \pi^-$ & (${}^\dagger$) & (${}^\dagger$) &
570 $\pm$ 40 \\ \hline
$\phi \to \pi^+ \pi^- \gamma$ & 6 $\cdot$ 10$^{-2}$ &
7 $\cdot$ 10$^{-2}$ & $<$ 30 \\
$\phi \to \pi^0 \pi^0 \gamma$ & 3 $\cdot$ 10$^{-2}$
 & 4 $\cdot$ 10$^{-2}$ & $<$ 4.4\\
$\phi \to \pi^0 \eta \gamma$ & 1.5 $\cdot$ 10$^{-2}$ &
1.8 $\cdot$ 10$^{-2}$ & $<$ 11 \\ \hline
$\phi \to \pi^+ \pi^- \pi^0$ & 65 & 60 & 106 $\pm$ 42 \\
\hline
\end{tabular} \end{center}
(${}^\dagger$) These data have been used as input.
\end{table}

\vspace*{0.5cm}

\noindent If instead, we use the value of $g_A\,=\, 0.76$
found above from the decay $\rho^+ \to \pi^+ \gamma$ then,
{}from $\phi \to \pi^0 \gamma$, we obtain

\be
\epsilon^2 \,=\, (4.8 \pm 0.6) \cdot 10^{-3} \,,\ee

\noindent and from $\phi \to \rho^+ \pi^-$

\be
\epsilon^2 \,=\, (6.3 \pm 0.5) \cdot 10^{-3} \,.\ee

\noindent Using the average of these two new results
we predict the decay rates for the $\phi$ vector that
are in the column Prediction 2. The only decay rate that
has been measured in Table \ref{table4}
is 1 $\sigma$ from our predictions.

\subsection{The decay $\pi^0 (\eta) \to \gamma \ell^+ \ell^-$}
\label{6.2} \indent

Here we shall study the process $\pi^0 (\eta) \to \gamma \gamma^*$
(and the related one $e^+ e^- \to \omega, \rho^0 \to \pi^0
\gamma$) in the limit of complete vector dominance,
{\it i.e.} assuming that the ${\cal O} (p^6)$ couplings are
saturated by the contribution coming from integrating out
the spin-1 particles as is the case for the ${\cal O}
(p^4)$ chiral Lagrangian couplings \cite{egpr}.
The amplitudes for these processes acquire a dependence on
the invariant mass of the off-shell photon from
chiral ${\cal O} (p^6)$ terms. Thus, one can define
a slope parameter as follows,

\vspace*{0.5cm}

\be \label{slope} \matrix{
\rho \equiv \left(
\frac{\dis 1}{\dis {\rm A} ( P \to \gamma \gamma^*)} \,
\, \frac{\dis {\rm d \hphantom{\, s^*}}}
{\dis {\rm d} \, s^*} \,\,{\rm A}
(P \to \gamma \gamma^*) \right)_{s^*=0}
\hfill \cr\cr } \ee

\vspace*{0.5cm}

\noindent which is independent of the pseudoscalar
meson $P$.
The slope $\rho$ has been measured in the $\pi^0
\to \gamma e^+ e^-$ transition \cite{pdg} with
the following result

\be \rho \,=\, (1.8 \pm 0.14) \,\, {\rm GeV}^{-2}\,. \ee

\vspace*{0.5cm}

\noindent
It has been also measured in the $\eta \to \gamma e^+ e^-$
transition \cite{pdg} and the result is

\be \rho \,=\, (1.4 \pm 0.2) \, {\rm GeV}^{-2}\,. \ee

\noindent
The weighted average from both results is

\be \label{sloex}
\rho \,=\, (1.68 \pm 0.2) \, {\rm GeV}^{-2}\,. \ee

\vspace*{0.5cm}

\noindent
Theoretically, the slope $\rho$ has two kinds of
contributions. One from chiral loops (see Figure
\ref{fig5}) and the other from the exchange
of vector mesons (see Figure \ref{fig6}).
The result we obtain in the ENJL model is the
following

\be \label{slopeth}
\rho \,=\, \frac{\dis 1}{\dis \Lambda_l^2} \,+\,
\frac{\dis 1}{\dis M_V^2} \, \left(
1 \,-\, \frac{\dis 2}{\dis 3} \,(1 \,-\, g_A) \right)\,, \ee

\vspace*{0.5cm}

\noindent where $\Lambda_l^2 \,=\, 3.57$ GeV$^2$
is the contribution from the chiral loops
which has been estimated in Ref. \cite{bij}.
The dependence in Eq. (\ref{slopeth}) on the coupling $g_A$
comes from the combination of vector couplings $f_V h_V$.
If we take for the parameters in Eq. (\ref{slopeth})
the ENJL predictions obtained with the choice in Eq. (\ref{param}),
{\it i.e.} $M_V \,=\, 0.8$ GeV and $g_A \,=\, 0.65$ \cite{bbr},
then we get

\be \rho \,=\, 1.48 \, {\rm GeV}^{-2}\,, \ee

\vspace*{0.5cm}

\noindent to be compared with the experimental result in
Eq. (\ref{sloex}). If instead, we take $g_A\,=\, 0.76$,
which is the other value we have been using
in the previous Section, we find

\be \rho \,=\, 1.59 \, {\rm GeV}^{-2}\,. \ee

\vspace*{0.5cm}

\noindent Both predictions are less than 1 $\sigma$
{}from the experimental result in Eq. (\ref{sloex}).

Next, we want to study the related process
$e^+ e^- \to \omega\,, \rho^0 \to
\pi^0 \gamma$. The Feynman diagrams for this
transition can be obtained from diagrams
in Figure \ref{fig5} and Figure \ref{fig6} by
inserting an $e^+ e^-$ pair
in the off-shell photon leg.
The cross section for this process is given by

\be \label{cross} \matrix{
\sigma_{e^+e^-\to\omega, \rho^0 \to
\pi^0\gamma} \, (s)\,=\, \frac{\dis \alpha^3}
{\dis 96 \, \pi^2 f_\pi^2} \, \left| \, {\dis \sum_
{V=\omega, \rho^0}} \,\left( C_l(\mu=M_V,s) \,+\,
1 \vphantom{\frac{\dis s}{\dis M_V^2}}
\right. \right. \hfill\cr\cr \left. \left.
\hphantom{\sigma_{e^+e^-\to\omega, \rho^0 \to
\pi^0\gamma} \, (s)\,=\,}\,+\,
\left(1\,-\, \frac{\dis 2}{\dis 3}
\left(1-g_A\right)\right) \,
\frac{\dis s}{\dis M_V^2 \,-\, s \,-\,i\,
\sqrt s \, \Gamma_V} \right) \, \right|^2
\hfill \cr\cr} \ee

\noindent where we have used the full expression for the
vector propagator as suggested in
Ref. \cite{bgp}. (In this Reference
this cross section was given in the HGS
model \cite{ban,fuji}.)
The function $C_l(\mu,s)$ in Eq. (\ref{cross})
comes from chiral loops \cite {bgp}
and has the following expression

\be \label{cl} \matrix{
C_l (\mu,s) \,=\, \frac{\dis 1}{\dis 48 \pi^2
\, f_\pi^2}\, \left[ s \, \ln \frac{\dis \mu^2}{\dis
m_\pi \, m_K} \, +\, \frac{\dis 5}{\dis 3}\, s \,+\,
2\, F(m_\pi^2,s)\,+\,2\,F(m_K^2,s) \right] \hfill
\cr\cr{\rm with} \hfill \cr\cr
F(m^2,s)\,=\, m^2\, \left(1\,-\,\frac{\dis x}{\dis 4}
\right)\, \left(1\,-\,\frac{\dis 4}{\dis x}\right)^{1/2}
\, \ln \frac{\dis \sqrt x \,+\, \sqrt{x-4}}{
\dis \sqrt x \,-\, \sqrt{x-4}} \,-\, 2\, m^2
\hfill\cr\cr {\rm for} \hspace*{0.2cm}
x\equiv\frac{\dis s}{\dis m^2}\,>\,4 \hspace*{0.5cm}
{\rm and} \hfill\cr\cr
F(m^2,s)\,=\, m^2\, \left(1\,-\,\frac{\dis x}{\dis 4}
\right)\, \left(\frac{\dis 4}{\dis x}\,-\,1\right)^{1/2}
\, \arctan \sqrt{\frac{\dis x}{\dis x\,-\,4}}\,-\, 2\, m^2
\hfill\cr\cr {\rm for} \hspace*{0.2cm} x\,\leq\,4\,.
\hfill \cr\cr} \ee

\vspace*{0.5cm}

\noindent The result we get for this cross section
at the $\omega$-mass peak using the set of parameters
in Eq. (\ref{param}) is

\be \label{nb1} \sigma ( e^+ e^- \to \omega, \rho \to
\pi^0 \gamma) {\dis |}_{s=M_\omega^2}\,=\,
94 \, {\rm nb} \, ,\ee

\vspace*{0.5cm}

\noindent to be compared with the experimental result
\cite{dol},

\be \sigma ( e^+ e^- \to \omega, \rho \to
\pi^0 \gamma) {\dis |}_{s=M_\omega^2}\,=\,
(152 \pm 13) \, {\rm nb}\,.\ee

\vspace*{0.5cm}

\noindent If instead,
we use the value for $g_A\,=\,0.76$ we find

\be \sigma ( e^+ e^- \to \omega, \rho \to
\pi^0 \gamma) {\dis |}_{s=M_\omega^2}\,=\,
113 \, {\rm nb}\,.\ee

\vspace*{0.5cm}

\noindent The last result is better (3 $\sigma$
{}from the experimental result).

\subsection{$a_1$-decays} \indent

In this Section we shall study the decays
corresponding to the axial-vector particle $a_1$.
These decays have been studied in Ref. \cite{ivan}
at the quark-loop level within a different ENJL model.
The results we find for these decays are presented
in Table \ref{table3} and the Feynman diagrams
in Figure \ref{fig7}.
This particle decays mainly into $\rho \pi$ and the
full width reported in Ref. \cite{pdg} ranges
between 350 and 450 MeV. The amplitudes that we obtain
for the transitions $a_1^+ \to \rho^+ \pi^0$ and
 $a_1^+ \to \rho^0 \pi^+$ are

\be \label{a1rho} \matrix{
A(a_1^+\to\rho^+(p_1)\pi^0(p_2))\,
=\,-\,A(a_1^+\to\rho^0(p_1)\pi^+(p_2))
\hfill\cr\cr \hphantom{A(a_1^+(P)\to
)}\,=\,i\,\frac{\dis 2}{\dis f_\pi}\,
\left[ A^{(2)}\,\left(\left(p_1 + p_2\right)^\mu
 \varepsilon^{*\nu}_{(a_1)}\,-
\,\left(p_1 + p_2\right)^\nu
\varepsilon^{*\mu}_{(a_1)}\right)\, \varepsilon^
{(\rho)}_\mu \,\, p_{2,\nu} \, \right. \hfill\cr\cr \left.
\hphantom{A(a_1^+(P)\to)\,=\,\frac{\dis 2}
{\dis f_\pi^2}} +\,
A^{(3)}\, \left( p_1^\mu \, \varepsilon^\nu_{(\rho)}\,-\,
p_1^\nu \, \varepsilon^\mu_{(\rho)}\right)\, \varepsilon^
{*(a_1)}_\mu \,\, p_{2,\nu} \right]
\hfill\cr\cr} \ee

\vspace*{0.5cm}

\noindent with the following corresponding decay rates

\be \label{a1rate} \matrix{
\Gamma(a_1^+\to\rho\pi)\,=\, \frac{\dis 1}{\dis 48 \,
\pi M_{a_1}}\, \lambda^{1/2}\left(1,
\frac{\dis m_\pi^2}{\dis M_{a_1}^2}, \frac{\dis
M_\rho^2}{\dis M_{a_1}^2}\right)\, {\dis \sum_{\rm pol}}
\left| A\right|^2 \,. \hfill\cr\cr} \ee

\vspace*{0.5cm}

We have also studied the direct decays into three pions
of the $a_1$ axial-vector.
The amplitude we get for the transition
$a_1^+ \to \pi^+ \pi^0 \pi^0$ is the following

\be \label{a13pia} \matrix{
A(a_1^+\to\pi^+(p_1)\pi^0(p_2)\pi^0(p_3))\,=\,
-\,i\, \frac{\dis 2 \sqrt 2}{\dis f_\pi^3}
\, \varepsilon^{*\mu}_{(a_1)}\,
\left[ \left(\gamma^{(1)}+\gamma^{(3)}\right)\,
\right.\hfill\cr\cr \left.
\hspace*{0.2cm} \times\,
\left((p_1 \cdot p_2)\, p_{3,\mu} \,+\, (p_1
 \cdot  p_3)\, p_{2,\mu} \right)\,+\,
\left(2\,\gamma^{(2)}\,-\,\gamma^{(1)}\,
+\,2\,\gamma^{(4)}\right)\,
(p_2  \cdot  p_3)\, p_{1,\mu} \right] \, ;
\hfill\cr\cr} \ee

\vspace*{0.5cm}

\noindent and for $a_1^+ \to \pi^- \pi^+ \pi^+$,

\be \label{a13pib} \matrix{
A(a_1^+\to\pi^-(p_1)\pi^+(p_2)\pi^+(p_3))\,=\,
-\,i\, \frac{\dis 2 \sqrt 2}{\dis f_\pi^3}
\, \varepsilon^{*\mu}_{(a_1)}\,
\left[ 2\, \left(\gamma^{(1)}+\gamma^{(3)}\right)\,
(p_2  \cdot  p_3)\, p_{1,\mu} \right. \hfill\cr\cr \left.
\hspace*{0.2cm} \,+\,
\left(2\, \gamma^{(2)}\,+\,\gamma^{(3)}\,
+\,2\, \gamma^{(4)}\right)\, \left(
(p_1  \cdot  p_2)\, p_{3,\mu}
\,+\, (p_1  \cdot  p_3)\, p_{2,\mu} \right)\right]\,.
\hfill\cr\cr} \ee

\vspace*{0.5cm}

\noindent The experimental information we have
 is just on $a_1 \to \left(\pi \pi\right)_S \pi$
decays, {\it i.e.} where
two of the pions are in an S-wave.
For this process
there is an educated guess in Ref. \cite{pdg}
 (the error is of the same order) for the experimental
 upper bound which is around 3 MeV.
If we assume that the reported partial width
for the decay $a_1 \to \left(\pi \pi
\right)_S \pi$ is only due to direct production
and not to a possible intermediate scalar particle,
{\it i.e.} $a_1 \to a_0 (f_0) \pi \to \left(\pi
\pi \right)_S \pi$ that would have been eventually
detected, then  we obtain the following
amplitudes. For the transition
$a_1^+ \to \left(\pi^+ \pi^0 \right)_S \pi^0$,

\be \label{a13pias} \matrix{
A(a_1^+\to\left(\pi^+(p_1)\pi^0(p_2)\right)_S
\pi^0(p_3))\,=\,\hfill\cr\cr
\hphantom{A(a_1^+\to\left(\pi^+(p_1)\pi^0(p_2)
\right)_S \,=\,}
-\,i\, \frac{\dis 2 \sqrt 2}{\dis f_\pi^3}
\, \left(\gamma^{(1)}\,+\,\gamma^{(3)}\right)\,
\, \left(\varepsilon^*_{(a_1)}  \cdot  p_3
\right) \, (p_1  \cdot  p_2) \, ; \hfill\cr\cr} \ee

\vspace*{0.5cm}

\noindent and for the $a_1^+ \to
\left(\pi^- \pi^+\right)_S \pi^+$,

\be \label{a13pibs} \matrix{
A(a_1^+\to\left(\pi^-(p_1)\pi^+(p_2)\right)_S
\pi^+(p_3))\,=\,\hfill\cr\cr
\hphantom{A(a_1^+\to \left(\pi^+(p_1)\right)}
-\,i\, \frac{\dis 2 \sqrt 2}{\dis f_\pi^3}
\, \left(2\,\gamma^{(2)}\,+\,\gamma^{(3)}\,
+\,2\,\gamma^{(4)}\right)
\, \left(\varepsilon^*_{(a_1)}  \cdot
p_3 \right) \, (p_1  \cdot  p_2) \, .\hfill\cr\cr} \ee

\vspace*{0.5cm}

The decay rate for $a_1^+ \to \pi^- \pi^+ \pi^+$
(for $a_1^+ \to \pi^+ \pi^0 \pi^0$ and
$a_1^+ \to \left(\pi \pi\right)_S \pi$
 are analogous) can be written as follows

\be \label{a13prate}
\Gamma(a_1^+ \to \pi^- \pi^+ \pi^+) \, = \,
\frac{\dis 1}{\dis 768 \, \pi^3 M_{a_1}} \, {\dis \int}
\, {\rm d} E_+ \, {\dis \int} \, {\rm d} E_- \,
{\dis \sum_{\rm pol}} \, \left|A \right|^2
\ee

\vspace*{0.5cm}

\noindent with $E_\pm\,=\, E_{\pi^+}\pm
E_{\pi^-}$ and the limits given by

\be \label{limitsc1} \matrix{
(E_+)_{\rm max} \, = \, M_{a_1} \, - \, m_\pi \,,
\hspace*{1.5cm}
(E_+)_{\rm min} \,=\, \frac{\dis M_{a_1}}{\dis 2} \,+\,
\frac{\dis 3 m_\pi^2}{\dis 2 M_{a_1}} \,,\hfill \cr\cr
(E_-)_{{\rm max}}
\,=\, \frac{\dis m_\pi^2 + M_{a_1}^2 - 2 x - 2
 \left(M_{\pi^+\pi^-}^2\right)_{{\rm min}}}
{\dis 2 M_{a_1}}\,, \hfill \cr\cr
(E_-)_{{\rm min}}
\,=\, \frac{\dis m_\pi^2 + M_{a_1}^2 - 2 x - 2
 \left(M_{\pi^+\pi^-}^2\right)_{{\rm max}}}
{\dis 2 M_{a_1}}\,, \hfill \cr\cr } \ee

\noindent with

\be \label{limitsc2} \matrix{
\left(M_{\pi^+\pi^-}^2\right)_{{\rm min}({\rm max})}  \,= \,
\frac{\dis 1}{\dis 8 (m_\pi^2 + x)} \, \left[
\left(M_{a_1}^2 - m_\pi^2\right)^2 \right. \hfill \cr\cr
\hspace*{0.1cm} \left. \,-\,
\left( \lambda^{1/2}(2(m_\pi^2+x), m_\pi^2, m_\pi^2)
+(-) \, \lambda^{1/2}(M_{a_1}^2, m_\pi^2, 2 (m_\pi^2+x))
\right)\right] \,, \hfill \cr\cr
x\equiv p_1 \cdot p_2
\,=\, M_{a_1} E_+ - \frac{\dis M_{a_1}^2 + m_\pi^2}
{\dis 2}\,. \hfill \cr\cr} \ee

\vspace*{0.5cm}

Another interesting decay is $a_1^+ \to \rho^+ \gamma$.
The amplitude we obtain is

\be \label{a1rhog} \matrix{
 A(a_1^+ \to \rho^+(p) \gamma(k)) \,=\, i\,|e|\,\frac{\dis 4}
{\dis 3} \,H\, \epsilon_{\mu \nu \alpha \beta}\,
\varepsilon^\mu_{(\rho)}\, \varepsilon^{* \nu}_{(a_1)}\,
k^\alpha \, \varepsilon^\beta_{(\gamma)}
\hfill\cr\cr } \ee

\vspace*{0.5cm}

\noindent and the corresponding decay rate

\be \label{a1rhpgr} \matrix{
 \Gamma(a_1^+\to \rho^+ \gamma) \,=\,\alpha \,
 \frac{\dis 2}{\dis 27}\,
 H^2 \, M_{a_1}\, \left(1\,-\, \frac{\dis
M_\rho^2}{\dis M_{a_1}^2}\right)^3\,
\left(1\,+\, \frac{\dis M_\rho^2}{\dis M_{a_1}^2}
\right) \, . \hfill\cr\cr} \ee

\vspace*{0.5cm}

Finally, we shall report on the decay
$a_1^+ \to \pi^+ \gamma$. The amplitude
we get for this process is

\be \label{a1pg} \matrix{
A(a_1^+\to \pi^+(p)\gamma(k))\,
=\, -\,i\, |e|\, \frac{\dis f_A\,-\, 2
\, \sqrt 2 \, \alpha_A}{\dis 2\, f_\pi}
\hfill\cr\cr \hphantom{A(a_1^+\to \pi^+(p))} \times \,
\, (\varepsilon_{(a_1)}^{*\mu} \, (p+k)^\nu\,-\,
\varepsilon_{(a_1)}^{*\nu} \, (p+k)^\mu)\,
(\varepsilon^{(\gamma)}_\mu k_\nu\,-\,
\varepsilon^{(\gamma)}_\nu k_\mu)
\hfill\cr\cr}\ee

\vspace*{0.5cm}

\noindent and the corresponding decay rate

\be \label{a1pgra} \matrix{
\Gamma(a_1^+\to\pi^+\gamma)\,=\,
\alpha \, \frac{\dis M_{a_1}^3}{\dis 24 \, f_\pi^2}
\, \left(f_A\,-\,2\sqrt 2 \, \alpha_A
\right)^2 \,\left(1\,-\, \frac{\dis m_\pi^2}{\dis
M_{a_1}^2}\right)^3 \, .\hfill\cr\cr} \ee

\vspace*{0.5cm}

The predictions for all these decays for the set
of input parameters in Eq. (\ref{param}) are the ones
in the column Prediction 1 in Table \ref{table3}.

\begin{table}[htb]
\caption{Partial widths in MeV corresponding to the
$a_1$-decays.}
\label{table3}
\begin{center}
\begin{tabular}{|l|c|c|c|}
\hline
 Process & Prediction 1 & Prediction 2 &Experiment\\
\hline
\hline
$a_1^+ \to \rho^+ \pi^0 \hphantom{\pi^0}$
& 130 & 178 &{\rm dominant}\\
$a_1^+ \to \rho^0 \pi^+ \hphantom{\pi^0}
$ &129&176&{\rm dominant}\\
$a_1^+ \to \pi^+ \pi^0 \pi^0$ & 0.8 & 1.9&--- \\
$a_1^+ \to \pi^- \pi^+ \pi^+$ & 1.0 & 2.5&---\\
$a_1^+ \to \left(\pi \pi\right)_S \pi$ & 0.6 & 1.4&$<$
3 (${}^*$)\\
$a_1^+ \to \rho^+ \gamma \hphantom{\pi^0}$
& 5 $\cdot$ 10$^{-3}$& 5 $\cdot$ 10$^{-3}$& ---\\
$a_1^+ \to \pi^+ \gamma \hphantom{\pi^0}$
&0.52 & 0.71 & 0.64 $\pm$ 0.25\\
\hline
\end{tabular} \end{center}
(${}^*$) This is an educated guess, the experimental
error is of the same order \cite{pdg}.
\vspace*{0.5cm} \end{table}

\noindent The result we obtain for the full
width of the $a_1$ axial-vector from the results in
the column Prediction 1 is $\Gamma_{\rm full}\,=$
261 MeV, to be compared
with the experimental value quoted at the beginning
of this Section (350 MeV $<$ $\left(\Gamma_{\rm full}
\right)_{\rm exp}$ $<$ 450 MeV).

In the column Prediction 2 we present the results
obtained using
the value  $g_A\,=\, 0.76$.
Given the poor experimental information we have on
$a_1$-decays, both sets of predictions in Table
\ref{table3} are quite good.
However, the full width for the $a_1$ axial-vector particle
we obtain from the results in the column
Prediction 2, $\Gamma_{\rm full}\,=$ 359 MeV,
 is better when compared with the experimental result
quoted above.

\subsection{$K_L \to \pi^0 \gamma^* \gamma^* \to \pi^0
e^+ e^-$} \indent

The decay $K_L \to \pi^0 e^+ e^-$ has received a great
deal of attention
as a possible candidate for observing CP-violation
effects (for a review see Ref. \cite{pr}).
The $K_L$ state consists mostly of a CP-odd state
$K_2$ with a small mixing of the CP-even state $K_1$,

\be \label{kl} \matrix{
K_L \simeq K_2\,+\, \epsilon\,K_1
\hfill\cr\cr} \ee

\vspace*{0.5cm}

\noindent where $|\epsilon| \simeq 2.26 \times 10^{-3}$
is the standard CP-violation parameter in $K \to \pi \pi$
decays. This decay is interesting because ``direct''
CP-violation in the $K_2$-decay amplitude
and ``indirect'' ($\epsilon$ effect) CP-violation
branching ratios are estimated to be of
the same order \cite{epr3}-\cite{fr}.
(Of the order of $10^{-11}$ \cite{ddg,fr}.)

In addition to these CP-violating decay modes there is
a two-photon exchange contribution to the transition
$K_L \to \pi^0 e^+ e^-$ which is CP-conserving.
In order to interpret future experimental measurements
of this decay it is crucial to elucidate if this decay mode may
compete or not with the CP-violating one-photon exchange
contributions. The two-photon exchange decay mode was studied in
Ref. \cite{epr3} at ${\cal O} (p^4)$
in the context of chiral perturbation theory
and was shown to be suppressed at this order
\cite{epr3,dono}. At ${\cal O} (p^6)$ there is
 a vector meson exchange diagram (shown in Figure \ref{fig8})
for the CP-conserving two-photon exchange contribution
which a priori could be potentially large \cite{seh}
and may compete with the one-photon CP-violating
decay modes. We can make a prediction for the
absortive ${\cal O} (p^6)$ vector meson dominance (VMD)
contribution to the process $K_L \to
\pi^0 \gamma^* \gamma^* \to \pi^0 e^+ e^-$ in the model
we are considering. To this VMD prediction one must add
further direct weak ${\cal O} (p^6)$ contributions, as was
pointed out in Ref. \cite{epr5}. In this Reference,
the whole ${\cal O} (p^6)$ contribution to $K_L \to \pi^0 \gamma^*
\gamma^* \to \pi^0 e^+ e^-$ was estimated by using the so-called ``weak
deformation'' model. Disregarding the ${\cal O} (p^4)$ helicity
suppressed contributions, these authors find

\be \label{brk} \matrix{
{\rm BR} \, \left(K_2 \to \pi^0 e^+ e^-\right)
{\dis |}_{\rm Abs} \simeq 4.4 \, a_V^2 \, 10^{-12}
\hfill\cr\cr}\ee

\noindent with

\be \left|a_V \right| \equiv \frac{\dis 512\, \pi^2\, h_V^2\, m_{K^0}^2}
{\dis 9 \, M_V^2} \ee

\vspace*{0.5cm}

\noindent where $h_V$ is the vector coupling introduced
in the Lagrangian in Eq. (\ref{47}) and the intermediate
resonance $V$ denotes the $\rho^0$ and $\omega$
vector particles.

With our determination of this coupling and using the
set of input parameters in Eq. (\ref{param}), we obtain
$h_V\,=$ 0.030 which leads to

\be \label{hv1} \matrix{
\left|a_V\right|\,=\, 0.21 \, .\hfill\cr\cr}\ee

\noindent
If we use the higher value for $g_A\,=\,0.76$,
which as we have seen is favoured by the overall data
on spin-1 particle decays, then we obtain
$h_V\,=$ 0.033 which leads to

\be \label{hv2} \matrix{
 \left|a_V\right|\,=\, 0.25 \, .\hfill\cr\cr}\ee

\noindent These values for $\left|a_V\right|$
are in good agreement with
the different phenomenological estimates \cite{fr,epr5}
made for  the  branching ratio in Eq. (\ref{brk}).

Recently, a measurement of the branching ratio for the
CP-conserving decay
$K_L \to \pi^0 \gamma \gamma$ has been reported by
 the NA31 experiment
at CERN \cite{na31}. This measurement gives the following
bounds at 90 $\%$ C.L.,

\be \matrix{ - \,0.32 \,<\,a_V\,< 0.19 \, .
\hfill\cr\cr } \ee

\section{Conclusions} \label{6} \indent

In this work we have studied in the context of the ENJL
cut-off model considered in Ref. \cite{bbr}
the anomalous and non-anomalous sectors of the
chiral Lagrangian for spin-1 particles
to  ${\cal O}(p^3)$, including terms with
two spin-1 particles.
In this model, spin-1 particles are introduced
not as gauge fields, in contrast to models in Refs.
\cite{ban}-\cite{meis}, but
as fields which transform homogeneously. This feature
gives rise to a formal symmetry of the QCD effective
action (see Section \ref{2}), first noticed in Ref. \cite{bbr},
that we use extensively troughout the text.
We have calculated 8 couplings in the anomalous sector
and 28 in the non-anomalous sector, both for terms involving
spin-1 particles. In particular, we have
given the dependence of these anomalous
and non-anomalous couplings on the axial-coupling $g_A$.
In this ENJL model, all the couplings we have calculated
here and those in the strong chiral Lagrangian describing
the interactions between Goldstone bosons \cite{bbr}
can be determined as functions
of just three parameters, namely $g_A$ , defined in
Eqs. (\ref{ga}) and (\ref{gad}), $x\,\equiv\,
M_Q^2/\Lambda^2_\chi$ and $M_Q$.
Taking the values for these three parameters
that fit the chiral
${\cal O}(p^2)$ and ${\cal O}
(p^4)$ low-energy couplings (see Ref. \cite{bbr})
we have given predictions for the following
low-energy
processes: $V \to \pi \gamma$, $V \to \pi \pi \pi$,
$V \to \pi \pi \gamma$, $V \to V' \gamma$, $V \to V'
\pi$, $A \to \pi \pi \pi$, $A \to \pi \gamma$,
$A \to V \pi$ and $A \to V \gamma$ at low orders
in the chiral expansion. The vector meson
dominance limit predictions for
$\pi^0 \to \gamma \gamma^*$ and
$K_L \to \pi^0 \gamma^* \gamma^* \to \pi^0 e^+ e^-$
processes have also been discussed.
The results are, in general, in good agreement with
the present experimental data taking into account that
they have been
obtained in the chiral limit. In addition, we make
predictions for decay rates which can be measured in future
low-energy experiments. We have also seen that the
available overall data on spin-1 decays
prefer a slightly higher value for the axial coupling $g_A$
than the one reported in Ref. \cite{bbr} obtained from a fit to
the low-energy data to chiral ${\cal O} (p^4)$.
In conclusion, we find that the ENJL model we have been considering
reproduces quite well a large variety of low-energy processes
where spin-1 particles are involved and the predictions we make
depend on just a small number of parameters (three) which
can be fixed from other low-energy transitions.

\section*{Acknowledgements} \indent

It is a pleasure to thank Eduardo de Rafael for
continuous encouragement and very helpful discussions.
I also wish to thank Hans Bijnens and Toni Pich
for useful conversations and Lars
H\"ornfeldt for his help with the algebraic manipulation
program STENSOR. This work has been
supported in part by CICYT, Spain, under Grant No.
AEN90-0040. I am indebted to the Spanish
Ministerio de  Educaci\'on y Ciencia for a fellowship.


\newpage \pagestyle{empty}
\begin{center}
\section*{FIGURE CAPTIONS} \end{center} \indent

\vspace*{1cm}

\noindent Figure \ref{fig1}: Diagrams contributing
to the process $V\to P\gamma$ ($P\to V\gamma$).

\vspace*{0.5cm}

\noindent Figure \ref{fig2}: Pion and kaon chiral loops
contribution to the process $\rho \to\pi\gamma$.

\vspace*{0.5cm}

\noindent Figure \ref{fig3}: Diagrams contributing to the
process $V\to \pi \pi \pi$.

\vspace*{0.5cm}

\noindent Figure \ref{fig4}: Diagrams contributing to the
process $V\to P_1 P_2 \gamma$.

\vspace*{0.5cm}

\noindent Figure \ref{fig5}: Pion and kaon chiral loops
contribution to the process $P^0 \to \gamma\gamma^*$.

\vspace*{0.5cm}

\noindent Figure \ref{fig6}: Vector-exchange contribution to the
process $P^0 \to \gamma\gamma^*$.

\vspace*{0.5cm}

\noindent Figure \ref{fig7}: Diagrams for the $a_1$-decays.

\vspace*{0.5cm}

\noindent Figure \ref{fig8}: Vector-exchange contribution
to the $K_L \to \pi^0 \gamma^* \gamma^* \to \pi^0 e^+ e^-$
transition.

\clearpage
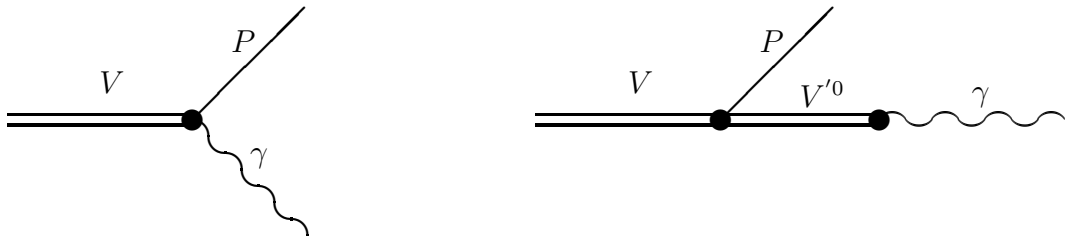
  \begin{figure}
  \begin{picture}(10000,18000) \THICKLINES
  \bigphotons
  \drawline\fermion[\E\REG](0,9200)[7000]
  \put(\pmidx,10000){$V$}
  \drawline\fermion[\E\REG](0,8800)[7000]
  \put(\fermionbackx,9000){\circle*{800}}
  \drawline\fermion[\NE\REG](\fermionbackx,9000)[6000]
  \global\advance\pmidx by -650
  \global\advance\pmidy by 430
  \put(\pmidx,\pmidy){$P$}
  \drawline\photon[\SE\REG](\fermionfrontx,9000)[7]
  \global\advance\pmidy by 600
  \put(\pmidx,\pmidy){$\gamma$}
  \drawline\fermion[\E\REG](20000,9200)[7000]
  \put(\pmidx,10000){$V$}
  \drawline\fermion[\E\REG](20000,8800)[7000]
  \put(\fermionbackx,9000){\circle*{800}}
  \drawline\fermion[\NE\REG](\fermionbackx,9000)[6000]
  \global\advance\pmidx by -650
  \global\advance\pmidy by 430
  \put(\pmidx,\pmidy){$P$}
  \drawline\fermion[\E\REG](\fermionfrontx,9200)[6000]
  \put(\pmidx,9500){$V^{'0}$}
  \drawline\fermion[\E\REG](\fermionfrontx,8800)[6000]
  \put(\fermionbackx,9000){\circle*{800}}
  \drawline\photon[\E\REG](\fermionbackx,9000)[7]
  \global\advance\pmidy by 700
  \put(\pmidx,\pmidy){$\gamma$}
  \end{picture}
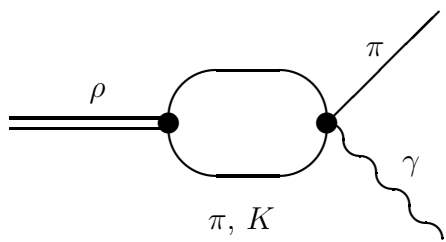
  \caption{} \label{fig1} \end{figure} \begin{figure}
  \begin{picture}(10000,18000) \THICKLINES
  \put(20000,9000){\oval(6000,4000)}
  \put(18500,5000){$\pi$, $K$}
  \pbackx=17000
  \drawline\fermion[\W\REG](\pbackx,9200)[6000]
  \put(\pmidx,10000){$\rho$}
  \drawline\fermion[\E\REG](\pbackx,8800)[6000]
  \put(\fermionbackx,9000){\circle*{800}}
  \pbackx=23000
  \put(\pbackx,9000){\circle*{800}}
  \drawline\fermion[\NE\REG](\pbackx,9000)[6000]
  \global\advance\pmidx by -650
  \global\advance\pmidy by 430
  \put(\pmidx,\pmidy){$\pi$}
  \drawline\photon[\SE\REG](\fermionfrontx,9000)[7]
  \global\advance\pmidx by 650
  \global\advance\pmidy by 430
  \put(\pmidx,\pmidy){$\gamma$}
  \end{picture} \caption{} \label{fig2} \end{figure}
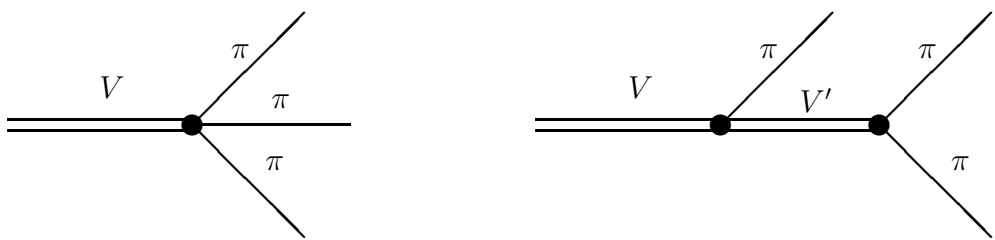
  \begin{figure}
  \begin{picture}(10000,18000) \THICKLINES
  \drawline\fermion[\E\REG](0,9200)[7000]
  \put(\pmidx,10000){$V$}
  \drawline\fermion[\E\REG](0,8800)[7000]
  \put(\fermionbackx,9000){\circle*{800}}
  \drawline\fermion[\NE\REG](\fermionbackx,9000)[6000]
  \global\advance\pmidx by -650
  \global\advance\pmidy by 430
  \put(\pmidx,\pmidy){$\pi$}
  \drawline\fermion[\E\REG](\fermionfrontx,9000)[6000]
  \global\advance\pmidy by 650
  \put(\pmidx,\pmidy){$\pi$}
  \drawline\fermion[\SE\REG](\fermionfrontx,9000)[6000]
  \global\advance\pmidx by 650
  \global\advance\pmidy by 430
  \put(\pmidx,\pmidy){$\pi$}
  \drawline\fermion[\E\REG](20000,9200)[7000]
  \put(\pmidx,10000){$V$}
  \drawline\fermion[\E\REG](20000,8800)[7000]
  \put(\fermionbackx,9000){\circle*{800}}
  \drawline\fermion[\NE\REG](\fermionbackx,9000)[6000]
  \global\advance\pmidx by -650
  \global\advance\pmidy by 430
  \put(\pmidx,\pmidy){$\pi$}
  \drawline\fermion[\E\REG](\fermionfrontx,9200)[6000]
  \put(\pmidx,9500){$V'$}
  \drawline\fermion[\E\REG](\fermionfrontx,8800)[6000]
  \put(\fermionbackx,9000){\circle*{800}}
  \drawline\fermion[\NE\REG](\fermionbackx,9000)[6000]
  \global\advance\pmidx by -650
  \global\advance\pmidy by 430
  \put(\pmidx,\pmidy){$\pi$}
  \drawline\fermion[\SE\REG](\fermionfrontx,9000)[6000]
  \global\advance\pmidx by 600
  \global\advance\pmidy by 430
  \put(\pmidx,\pmidy){$\pi$}
  \end{picture} \caption{} \label{fig3} \end{figure}
  \clearpage
  \begin{figure}
  \begin{picture}(10000,18000) \THICKLINES
  \drawline\fermion[\E\REG](0,9200)[7000]
  \put(\pmidx,10000){$V$}
  \drawline\fermion[\E\REG](0,8800)[7000]
  \put(\fermionbackx,9000){\circle*{800}}
  \drawline\fermion[\NE\REG](\fermionbackx,9000)[6000]
  \global\advance\pmidx by -850
  \global\advance\pmidy by 430
  \put(\pmidx,\pmidy){$P_1$}
  \drawline\photon[\E\REG](\fermionfrontx,9000)[7]
  \global\advance\pmidy by 650
  \put(\pmidx,\pmidy){$\gamma$}
  \drawline\fermion[\SE\REG](\photonfrontx,9000)[6000]
  \global\advance\pmidx by 700
  \global\advance\pmidy by 40
  \put(\pmidx,\pmidy){$P_2$}
  \drawline\fermion[\E\REG](20000,9200)[7000]
  \put(\pmidx,10000){$V$}
  \drawline\fermion[\E\REG](20000,8800)[7000]
  \put(\fermionbackx,9000){\circle*{800}}
  \drawline\fermion[\NE\REG](\fermionbackx,9000)[6000]
  \global\advance\pmidx by -2000
  \global\advance\pmidy by 430
  \put(\pmidx,\pmidy){$P_{1(2)}$}
  \drawline\fermion[\E\REG](\fermionfrontx,9200)[7000]
  \global\advance\pmidx by -2000
  \put(\pmidx,7000){$V'(V'')$}
  \drawline\fermion[\E\REG](\fermionfrontx,8800)[7000]
  \put(\fermionbackx,9000){\circle*{800}}
  \drawline\fermion[\NE\REG](\fermionbackx,9000)[6000]
  \global\advance\pmidx by -2000
  \global\advance\pmidy by 430
  \put(\pmidx,\pmidy){$P_{2(1)}$}
  \drawline\photon[\SE\REG](\fermionfrontx,9000)[7]
  \global\advance\pmidx by 600
  \global\advance\pmidy by 430
  \put(\pmidx,\pmidy){$\gamma$}
  \end{picture}
  \end{figure}
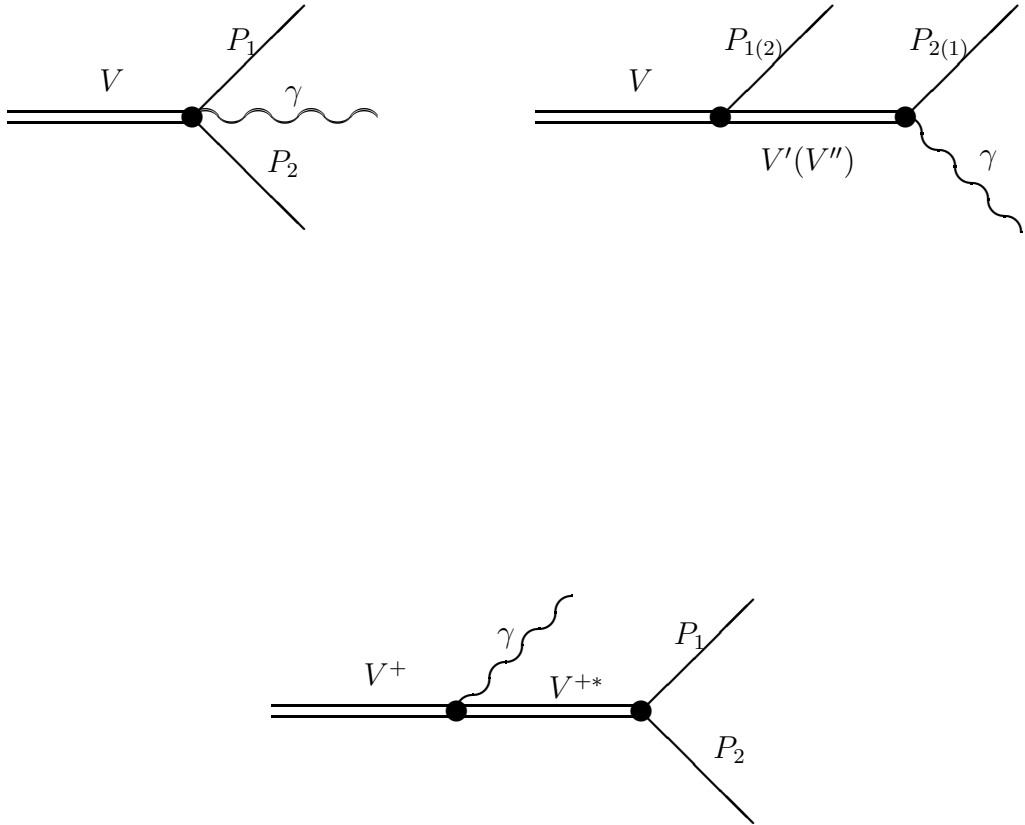
  \begin{figure}
  \begin{picture}(10000,18000) \THICKLINES
  \drawline\fermion[\E\REG](10000,9200)[7000]
  \put(\pmidx,10000){$V^+$}
  \drawline\fermion[\E\REG](10000,8800)[7000]
  \put(\fermionbackx,9000){\circle*{800}}
  \drawline\photon[\NE\REG](\fermionbackx,9000)[7]
  \global\advance\pmidx by -650
  \global\advance\pmidy by 430
  \put(\pmidx,\pmidy){$\gamma$}
  \drawline\fermion[\E\REG](\photonfrontx,9200)[7000]
  \put(\pmidx,9500){$V^{+*}$}
  \drawline\fermion[\E\REG](\fermionfrontx,8800)[7000]
  \put(\fermionbackx,9000){\circle*{800}}
  \drawline\fermion[\NE\REG](\fermionbackx,9000)[6000]
  \global\advance\pmidx by -900
  \global\advance\pmidy by 430
  \put(\pmidx,\pmidy){$P_1$}
  \drawline\fermion[\SE\REG](\fermionfrontx,9000)[6000]
  \global\advance\pmidx by 600
  \global\advance\pmidy by 350
  \put(\pmidx,\pmidy){$P_2$}
  \end{picture} \vspace*{1cm} \caption{} \label{fig4}
  \vspace*{6cm} \end{figure} \clearpage
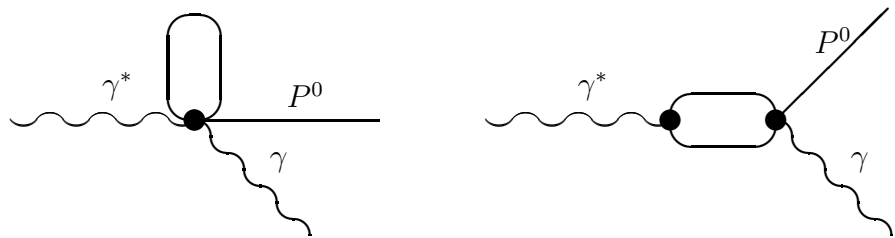
  \begin{figure}
  \begin{picture}(10000,18000) \THICKLINES
  \put(10000,11000){\oval(2000,4000)}
  \pbacky=9000 \pbackx=10000
  \drawline\photon[\W\REG](\pbackx,\pbacky)[7]
  \put(\pmidx,10000){$\gamma^*$}
  \drawline\fermion[\E\REG](\pfrontx,\pfronty)[7000]
  \global\advance\pmidy by 500
  \put(\pmidx,\pmidy){$P^0$}
  \put(\pfrontx,\pfronty){\circle*{800}}
  \drawline\photon[\SE\REG](\pfrontx,\pfronty)[7]
  \global\advance\pmidx by 650
  \global\advance\pmidy by 430
  \put(\pmidx,\pmidy){$\gamma$}
  \put(\pfrontx,\pfronty){\circle*{800}}
  \put(30000,9000){\oval(4000,2000)}
  \pbacky=9000 \pbackx=28000
  \drawline\photon[\W\REG](\pbackx,\pbacky)[7]
  \put(\pmidx,10000){$\gamma^*$}
  \put(\pfrontx,\pfronty){\circle*{800}}
  \pbacky=9000 \pbackx=32000
  \drawline\fermion[\NE\REG](\pbackx,\pbacky)[6000]
  \global\advance\pmidx by -650
  \global\advance\pmidy by 430
  \put(\pmidx,\pmidy){$P^0$}
  \drawline\photon[\SE\REG](\pfrontx,\pfronty)[7]
  \global\advance\pmidx by 650
  \global\advance\pmidy by 430
  \put(\pmidx,\pmidy){$\gamma$}
  \put(\pfrontx,\pfronty){\circle*{800}}
  \end{picture}
  \caption{} \label{fig5} \end{figure}
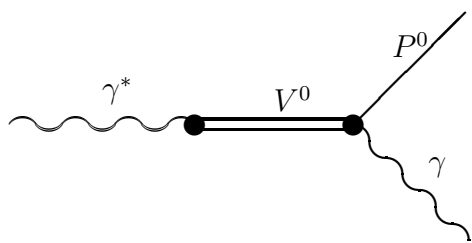
  \begin{figure}
  \begin{picture}(10000,18000) \THICKLINES
  \drawline\photon[\E\REG](10000,9000)[7]
  \put(\pmidx,10000){$\gamma^*$}
  \put(\photonbackx,9000){\circle*{800}}
  \drawline\fermion[\E\REG](\photonbackx,9200)[6000]
  \put(\pmidx,9500){$V^0$}
  \drawline\fermion[\E\REG](\fermionfrontx,8800)[6000]
  \put(\fermionbackx,9000){\circle*{800}}
  \drawline\fermion[\NE\REG](\fermionbackx,9000)[6000]
  \global\advance\pmidx by -650
  \global\advance\pmidy by 430
  \put(\pmidx,\pmidy){$P^0$}
  \drawline\photon[\SE\REG](\fermionfrontx,9000)[7]
  \global\advance\pmidx by 650
  \global\advance\pmidy by 430
  \put(\pmidx,\pmidy){$\gamma$}
  \end{picture} \caption{} \label{fig6}
  \end{figure}\clearpage
  \begin{figure}
  \begin{picture}(10000,18000) \THICKLINES
  \drawline\fermion[\E\REG](0,10200)[7000]
  \put(\pmidx,11000){$a_1$}
  \drawline\fermion[\E\REG](0,9800)[7000]
  \put(\fermionbackx,10000){\circle*{800}}
  \drawline\fermion[\NE\REG](\fermionbackx,10000)[6000]
  \global\advance\pmidx by -650
  \global\advance\pmidy by 430
  \put(\pmidx,\pmidy){$\pi$}
  \drawline\fermion[\E\REG](\fermionfrontx,10000)[6000]
  \global\advance\pmidy by 650
  \put(\pmidx,\pmidy){$\pi$}
  \drawline\fermion[\SE\REG](\fermionfrontx,10000)[6000]
  \global\advance\pmidx by 650
  \global\advance\pmidy by 430
  \put(\pmidx,\pmidy){$\pi$}
  \drawline\fermion[\E\REG](26000,10200)[7000]
  \put(\pmidx,11000){$a_1$}
  \drawline\fermion[\E\REG](26000,9800)[7000]
  \put(\fermionbackx,10000){\circle*{800}}
  \drawline\fermion[\NE\REG](\fermionbackx,10000)[6000]
  \global\advance\pmidx by -650
  \global\advance\pmidy by 430
  \put(\pmidx,\pmidy){$\pi$}
  \drawline\fermion[\SE\REG](\fermionfrontx,10250)[6300]
  \global\advance\pmidx by 600
  \global\advance\pmidy by 430
  \put(\pmidx,\pmidy){$\rho$}
  \drawline\fermion[\SE\REG](\fermionfrontx,9750)[5950]
  \end{picture} \end{figure}
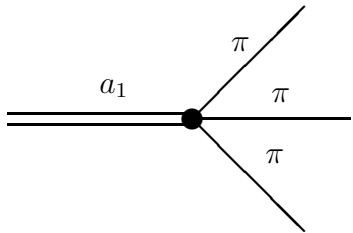
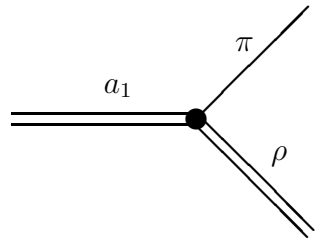
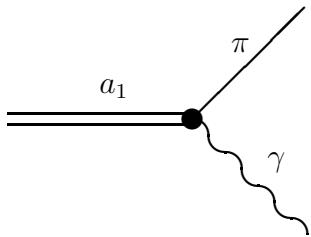
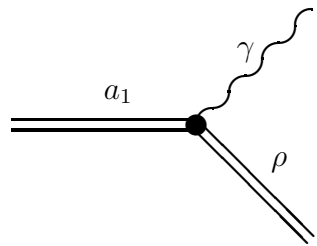
  \begin{figure}
  \begin{picture}(10000,18000) \THICKLINES
  \drawline\fermion[\E\REG](0,9200)[7000]
  \put(\pmidx,10000){$a_1$}
  \drawline\fermion[\E\REG](0,8800)[7000]
  \put(\fermionbackx,9000){\circle*{800}}
  \drawline\fermion[\NE\REG](\fermionbackx,9000)[6000]
  \global\advance\pmidx by -650
  \global\advance\pmidy by 430
  \put(\pmidx,\pmidy){$\pi$}
  \drawline\photon[\SE\REG](\fermionfrontx,9000)[7]
  \global\advance\pmidx by 650
  \global\advance\pmidy by 430
  \put(\pmidx,\pmidy){$\gamma$}
  \drawline\fermion[\E\REG](26000,10200)[7000]
  \put(\pmidx,11000){$a_1$}
  \drawline\fermion[\E\REG](26000,9800)[7000]
  \put(\fermionbackx,10000){\circle*{800}}
  \drawline\photon[\NE\REG](\fermionbackx,10000)[7]
  \global\advance\pmidx by -650
  \global\advance\pmidy by 430
  \put(\pmidx,\pmidy){$\gamma$}
  \drawline\fermion[\SE\REG](\fermionbackx,10250)[6300]
  \global\advance\pmidx by 600
  \global\advance\pmidy by 430
  \put(\pmidx,\pmidy){$\rho$}
  \drawline\fermion[\SE\REG](\fermionfrontx,9750)[5950]
   \end{picture} \vspace*{1cm} \caption{} \label{fig7}
  \vspace*{6cm} \end{figure} \clearpage
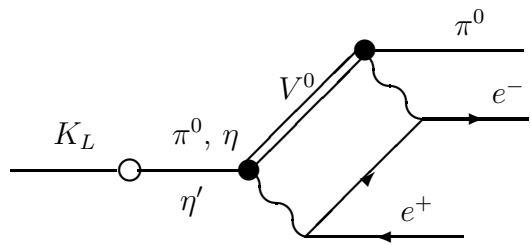
  \begin{figure}
  \begin{picture}(10000,18000) \THICKLINES
  \drawline\fermion[\E\REG](8000,10000)[4000]
  \global\advance\pmidx by -400
  \put(\pmidx,11000){$K_L$}
  \global\advance\fermionbackx by 500
  \put(\fermionbackx,10000){\circle{800}}
  \global\advance\fermionbackx by 300
  \drawline\fermion[\E\REG](\fermionbackx,10000)[4000]
  \global\advance\pmidx by -700
  \put(\pmidx,11000){$\pi^0$, $\eta$}
  \global\advance\pmidx by 300
  \put(\pmidx,8500){$\eta'$}
  \global\advance\fermionbackx by 200
  \put(\fermionbackx,10000){\circle*{800}}
  \global\advance\fermionbackx by -200
  \drawline\photon[\SE\REG](\fermionbackx,10000)[4]
  \drawline\fermion[\NE\REG](\fermionbackx,10250)[6300]
  \global\advance\pmidx by -900
  \global\advance\pmidy by 230
  \put(\pmidx,\pmidy){$V^0$}
  \drawline\fermion[\NE\REG](\fermionfrontx,9750)[6500]
  \global\advance\fermionbacky by 200
  \put(\fermionbackx,\fermionbacky){\circle*{800}}
  \drawline\fermion[\E\REG](\fermionbackx,\fermionbacky)[6000]
  \global\advance\pmidx by 400
  \global\advance\pmidy by 600
  \put(\pmidx,\pmidy){$\pi^0$}
  \global\advance\fermionfrontx by -300
  \global\advance\fermionfronty by -100
  \drawline\photon[\SE\REG](\fermionfrontx,\fermionfronty)[4]
  \drawline\fermion[\E\REG](\photonbackx,\photonbacky)[4000]
  \drawarrow[\E\ATBASE](\pmidx,\pmidy)
  \global\advance\pmidx by 600
  \global\advance\pmidy by 600
  \put(\pmidx,\pmidy){$e^-$}
  \drawline\fermion[\SW\REG](\photonbackx,\photonbacky)[6300]
  \global\advance\pmidx by 180
  \drawarrow[\NE\ATBASE](\pmidx,\pmidy)
  \drawline\fermion[\E\REG](\fermionbackx,\fermionbacky)[6000]
  \drawarrow[\W\ATBASE](\pmidx,\pmidy)
  \global\advance\pmidx by 600
  \global\advance\pmidy by 600
  \put(\pmidx,\pmidy){$e^+$}
   \end{picture} \vspace*{1cm} \caption{} \label{fig8}
  \vspace*{6cm} \end{figure} } \end{document}